\documentclass[pdflatex,sn-mathphys-num,iicol]{sn-jnl}

\usepackage{graphicx}
\usepackage{booktabs}
\usepackage{amsmath,amssymb,amsfonts}
\usepackage{bm}

\begin{document}


\title[Long-range voter model]{\textbf{Exact solution of the long-range voter model in arbitrary dimensions: correlations and effective-dimensional scaling}}





\author[1]{\fnm{Fabrizio} \sur{Rippa}}

\author[1]{\fnm{Jacopo Alexander} \sur{Garofalo}}

\author[2]{\fnm{Giuseppe} \sur{Petrillo}}

\author[1]{\fnm{Eugenio} \sur{Lippiello}}

\makeatletter
\def\@authfrstarg{1}
\makeatother


\affil[1]{
\orgdiv{Department of Mathematics and Physics},
\orgname{University of Campania ``L. Vanvitelli''},
\orgaddress{
\city{Caserta},
\country{Italy}
}
}

\affil[2]{
\orgname{CNRS, ENS de Lyon, LPENSL, UMR5672},
\orgaddress{
\postcode{69342},
\city{Lyon Cedex 07},
\country{France}
}
}


\abstract{
Long-range interactions can profoundly alter collective ordering and critical behavior, yet exact analytical results for non-equilibrium systems with such interactions remain scarce. We obtain an exact analytical solution for the spatial correlation function of the voter model with long-range interactions in arbitrary spatial dimension.
Starting from the evolution equation for the Fourier modes, we identify the long-wavelength dispersion generated by the interaction kernel and derive the resulting coarsening and stationary scaling regimes. The solution yields the critical condition separating these regimes, the characteristic coarsening length, and the asymptotic decay of stationary correlations. These results establish an exact effective-dimensional correspondence: the long-range model in dimension \(d\), with interaction exponent \(\beta\), exhibits the same scaling exponents as a short-range voter model in the effective dimension \(d_{\mathrm{eff}}=2d/\beta\). The correspondence holds across the coarsening, marginal, and stationary regimes, while the associated scaling functions need not coincide.} 



\maketitle


\section{Introduction}
\label{sec:introduction}
The voter model is a paradigmatic stochastic model of collective ordering, introduced independently by Clifford and Sudbury~\cite{CliffordSudbury1973} and by Holley and Liggett~\cite{HolleyLiggett1975}. It provides a minimal setting for ordering with absorbing states and anomalous coarsening, while in sociophysics it represents neutral imitation among interacting agents~\cite{Liggett1999,KrapivskyRednerBenNaim2010,CastellanoFortunatoLoreto2009,SenChakrabarti2014,starnini2026opinion}. Despite its simple microscopic rule, the model displays nontrivial consensus and coexistence phenomena and remains sufficiently tractable for analytical approaches~\cite{HenkelStoimenov2026}. Its relevance to opinion dynamics has also motivated comparisons with electoral data~\cite{FernandezGraciaEtAl2014,BrahaDeAguiar2017,SanoHisakadoMori2017}.

A large number of extensions have been proposed to incorporate nonlinear influence, group interactions, committed agents, and memory effects~\cite{redner2019reality}. Representative examples include the $q$- and $p$-voter models~\cite{CastellanoMunozPastorSatorras2009,CorberiDelloRussoSmaldone2024,jedrzejewski2015oscillating}, Sznajd--Weron dynamics~\cite{SznajdWeronSznajd2000,SznajdWeron2005}, voter models with zealots~\cite{Mobilia2003,MobiliaPetersenRedner2007}, and persistent voter dynamics~\cite{LatoskiDantasArenzon2024}. Unlike kinetic Ising ordering, voter dynamics is not driven by an energy functional or interfacial surface tension; except for the one-dimensional nearest-neighbor case, which is equivalent to zero-temperature Glauber dynamics, the two processes belong to different dynamical classes~\cite{DornicChateChaveHinrichsen2001,Glauber1963,KrapivskyRednerBenNaim2010}.

Long-range interactions provide a natural way to model nonlocal influence, including media exposure, weak ties, mobility, and geographically distant contacts~\cite{QuattrociocchiCaldarelliScala2014,ColaioriCastellano2015,Granovetter1973,ZhaoWuXu2010,ParkBlumenstockMacy2018}. They are also of fundamental interest because they modify phase-ordering and scaling properties. In long-range Ising systems the resulting nonlocal energetic couplings generally complicate both analysis and simulations~\cite{CorberiLippielloPoliti2019a,CorberiLippielloPoliti2019b,AgrawalEtAl2021,CorberiEtAl2021,CorberiKumarLippielloPoliti2023,AgrawalCorberiLippielloPuri2025}. In voter dynamics, by contrast, nonlocality enters through the sampling of copying events rather than through an energy landscape. This distinction underlies the crossover between voter-like and Ising-like long-range ordering discussed in recent works~\cite{CorberiLippielloPolitiSmaldone2026,CorberiDelloRussoSmaldone2024}.

Long-range voter models with copying probability $P(r)\sim r^{-d-\sigma}$ have been studied analytically, assuming dynamical scaling, in one, two, and three dimensions~\cite{CorberiCastellano2024,CorberiSmaldone2024,CorberiDelloRussoSmaldone2024b}. These studies identify a crossover at $\sigma_c=2$: for $0<\sigma<\sigma_c$ the nonanalytic long-range contribution controls the infrared dynamics, whereas for $\sigma>\sigma_c$ the leading bulk behavior is diffusive, although algebraic long-distance corrections remain. 

Here we investigate the long-range voter dynamics in arbitrary spatial dimension $d$. At each update, a selected spin copies another spin at distance $r$ drawn from the algebraic distribution $P(r)\sim r^{-d-\sigma}$. Away from $\sigma=2$, the leading infrared dynamics is controlled by $\beta=\min\{\sigma,2\}$. The competition between $d$ and $\beta$ determines whether the system coarsens indefinitely ($d\leq\beta$) or approaches a nontrivial stationary state ($d>\beta$). The characteristic length grows as $t^{1/\beta}$, while for $\sigma>2$ the algebraic interaction still produces subleading spatial tails. We finally show that the bulk infrared exponents can be organized through an effective dimension $d_{\rm eff}=2d/\beta$~\cite{Angelini2014,Defenu2015,Solfanelli2024,pagni2026one}, which maps the coarsening threshold and the principal scaling exponents onto their short-range voter counterparts. 

\section{The model}
\label{sec:model}

The model consists of $N=L^d$ voters on a $d$-dimensional hypercubic lattice of linear size $L$. Each site carries a binary variable $S_i=\pm1$.

During an elementary update, a site $i$ is selected at random and copies the state of a site $j$ sampled according to an algebraically decaying distance distribution. A Monte Carlo step consists of $N$ elementary updates.

To define the long-range process on a lattice, we group the sites according to their proximity to a given site. We denote by $p$ the proximity index, by $l_p$ the corresponding distance, and by $n_p$ the number of lattice sites belonging to the $p$-th proximity shell. Thus, for instance, $p=1$ identifies the nearest-neighbors shell, for which $l_1=1$ and $n_1=2d$ on a hypercubic lattice. 

The probability associated with an individual site at distance $l_p$ is taken as 
\begin{equation}
    P(l_p)=\frac{1}{Z_d(\sigma)}\,l_p^{-d-\sigma}\,,
\end{equation}
where $\sigma>0$ controls the range of the interaction and
\begin{equation}
    Z_d(\sigma)=\sum_p n_p\,l_p^{-d-\sigma}
\end{equation}
is the normalization constant.

The spin-flip transition rate for the $i$-th spin is
\begin{equation}
    w(S_i)=\frac{1}{2N}\,\sum_pP(l_p)\,\sum_{k=1}^{n_p}(1-S_i\,S_k)\,.
    \label{eq. LR transition rate}
\end{equation}
Here the inner sum runs over the $n_p$ sites belonging to the $p$-th shell around site $i$.

We characterize the dynamics through the equal-time correlation function $C(\boldsymbol r,t)=\langle S_i(t)S_j(t)\rangle$, with $\boldsymbol r=\boldsymbol r_j-\boldsymbol r_i$. For translationally invariant initial conditions, the evolution equation, with time measured in Monte Carlo steps, following from Eq.~(\ref{eq. LR transition rate}) takes the exact lattice form~\cite{Glauber1963}
\begin{equation}
\begin{aligned}
\frac{\partial C(\boldsymbol r,t)}{\partial t}
={}&-2C(\boldsymbol r,t)
+2\sum_p P(l_p)\\
&\times\sum_{k=1}^{n_p} C(\boldsymbol r-\boldsymbol{\ell}_{p,k},t)
+v(t)\,\delta_{\boldsymbol r,0}\,.
\end{aligned}
    \label{eq. exact discrete}
\end{equation}
Here \(\boldsymbol{\ell}_{p,k}\) denotes the displacement vector connecting site \(i\) to the \(k\)-th site in the \(p\)-th proximity shell, with \(|\boldsymbol{\ell}_{p,k}|=l_p\).  
The term $v(t)$ is a Lagrange multiplier introduced to impose the constraint
\begin{equation}
    C(0,t)=1\,,\qquad \forall t\,.
    \label{eq. lagrange constraint}
\end{equation}
This source-term construction follows Ref.~\cite{GodrecheLuck2000}. 
We introduce the discrete Fourier transform
\begin{equation}
C^F(\boldsymbol q,t)
=
\sum_{\boldsymbol r}
e^{-i\boldsymbol q\cdot\boldsymbol r}
C(\boldsymbol r,t),
\label{eq. Fourier transform}
\end{equation}
with inverse transform
\begin{equation}
C(\boldsymbol r,t)
=
\frac{1}{N}\sum_{\boldsymbol q}
e^{i\boldsymbol q\cdot\boldsymbol r}
C^F(\boldsymbol q,t),
\end{equation}
where $\boldsymbol q=(2\pi/L)\boldsymbol n$, with $\boldsymbol{n}\in\{0,1,\ldots,L-1\}^d$, runs over
the allowed wavevectors of the lattice.

In Fourier space, the exact lattice evolution
equation~(\ref{eq. exact discrete}) becomes (see Appendix~\ref{app:kernel})
\begin{equation}
\frac{\partial C^F(\boldsymbol q,t)}{\partial t}
=
-\Lambda_{\sigma}^{(d)}(\boldsymbol q)
C^F(\boldsymbol q,t)+v(t),
\label{eq. fourier evolution}
\end{equation}
where the lattice generator is
\begin{equation}
\Lambda_{\sigma}^{(d)}(\boldsymbol q)
=
2\sum_p P(l_p)\sum_{k=1}^{n_p}
\left[
1-\cos(\boldsymbol q\cdot\boldsymbol\ell_{p,k})
\right].
\label{eq. Lambda}
\end{equation}

Equation~(\ref{eq. fourier evolution}), together with
Eq.~(\ref{eq. Lambda}), provides an exact description
of the lattice dynamics.

As shown in Appendix~\ref{app:kernel}, the
long-wavelength expansion of the lattice generator
contains an analytic quadratic contribution and
a nonanalytic term arising from the algebraic
interaction kernel. Away from marginal values,
these contributions take the form

\begin{equation}
\Lambda_{\sigma}^{(d)}(\boldsymbol q)
=
\begin{cases}
D_{\sigma}^{(d)}q^\sigma+o(q^\sigma),
& 0<\sigma<2,\\[4pt]
B_{\sigma}^{(d)}q^2+
D_{\sigma}^{(d)}q^\sigma+o(q^\sigma),
& 2<\sigma<4,
\end{cases}
\label{eq. Lambda expansion}
\end{equation}
Here \(q=|\boldsymbol q|\). For \(0<\sigma<2\), the leading contribution is nonanalytic, whereas for \(\sigma>2\) the quadratic term dominates. The algebraic interaction kernel also generates a subleading nonanalytic contribution for \(\sigma>2\), which remains relevant to the large-distance behavior of the correlations. For \(\sigma\geq4\), additional analytic terms may precede this contribution, while logarithmic corrections arise at marginal even values of \(\sigma\). The detailed derivation is presented in Appendix~\ref{app:kernel}.

The leading long-wavelength behavior is therefore
isotropic. In particular, for rotationally invariant
initial conditions, the
Fourier-space correlation function $C^F(\boldsymbol q,t)$ depends on
$\boldsymbol q$ only through its magnitude $q$,
up to subleading corrections. Consequently,
rotational invariance also emerges in real space,
where the leading scaling behavior of
$C(\boldsymbol r,t)$ depends only on the
separation $r=|\boldsymbol r|$.

The formal solution of Eq.~(\ref{eq. fourier evolution})
is therefore
\begin{equation}
\begin{aligned}
C^F(q,t)={}&e^{-\Lambda_{\sigma}^{(d)}(q)t}C^F(q,0)\\
&+\int_0^t dt'\,v(t')\,
 e^{-\Lambda_{\sigma}^{(d)}(q)(t-t')}\,.
\end{aligned}
\end{equation}
For a fully disordered initial condition
\begin{equation}
    C(\boldsymbol r,0)=\delta^{(d)}(\boldsymbol r)\,,\qquad C^F(\boldsymbol q,0)=1\,,
\end{equation}
and hence
\begin{equation}
\begin{aligned}
C^F(q,t)={}&e^{-\Lambda_{\sigma}^{(d)}(q)t}\\
&+\int_0^t dt'\,v(t')\,
 e^{-\Lambda_{\sigma}^{(d)}(q)(t-t')}\,.
\end{aligned}
    \label{eq. formal Fourier space solution}
\end{equation}


For later convenience, away from the marginal value $\sigma=2$, we introduce the infrared exponent
\begin{equation}
    \beta=
    \begin{cases}
        \sigma\,, & 0<\sigma<2\,,\\[2mm]
        2\,, & \sigma>2\,,
    \end{cases}
    \label{eq. beta definition}
\end{equation}
together with the coefficient of the leading small-$q$ contribution,
\begin{equation}
    \mathcal{D}_{\beta}^{(d)}=
    \begin{cases}
        D_{\sigma}^{(d)}\,, & 0<\sigma<2\,,\\[2mm]
        B_{\sigma}^{(d)}\,, & \sigma>2\,.
    \end{cases}
    \label{eq. Dbeta definition}
\end{equation}
At the marginal value \(\sigma=2\), the second moment of the interaction kernel diverges logarithmically. As a consequence (see Appendix~\ref{app:kernel}), the leading long-wavelength behavior of the lattice generator acquires a logarithmic correction,
\begin{equation*}
\Lambda_2^{(d)}( q)\sim q^2\log(1/q),    
\end{equation*}
The marginal case therefore requires a separate treatment and is not considered in the following.

The leading infrared behavior of the generator can then be written in the unified form
\begin{equation}
    \Lambda_{\sigma}^{(d)}(q)\simeq \mathcal{D}_{\beta}^{(d)}\,|q|^\beta\,,\qquad q\rightarrow0\,.
    \label{eq. unified infrared Lambda}
\end{equation}
The exponent $\beta$ controls the leading bulk behavior of the long-range dynamics.


\section{Stationary states}
\label{sec:stationary}

In this section we investigate the existence and spatial structure of stationary states in the purely long-range voter dynamics. Throughout this section the thermodynamic limit is assumed. The dynamical process through which they are approached will be discussed separately in Sec.~\ref{sec:approach stationary}.

In the stationary state, $C^F(q,t)\xrightarrow{t\rightarrow\infty} C^F_{\rm st}(q)$ and the Lagrange multiplier approaches a time-independent value $v_{\rm st}$. Eq.~\eqref{eq. fourier evolution} then gives
\begin{equation}
    C_{\rm st}^F(q)=\frac{v_{\rm st}}{\Lambda_{\sigma}^{(d)}(q)}\,.
    \label{eq. st correlation}
\end{equation}
The value of $v_{\rm st}$ follows from the normalization condition (Eq.~\eqref{eq. lagrange constraint}). In Fourier space,
\begin{equation}
\begin{aligned}
1&=\int_{\rm BZ}\frac{d^dq}{(2\pi)^d}\,C^F_{\rm st}(q)\\
 &=v_{\rm st}\int_{\rm BZ}\frac{d^dq}{(2\pi)^d}\,
  \frac{1}{\Lambda_{\sigma}^{(d)}(q)}\,.
\end{aligned}
    \label{eq. IR st integral}
\end{equation}
and therefore
\begin{equation}
    v_{\rm st}
    =
    \left[
    \int_{\rm BZ}\frac{d^dq}{(2\pi)^d}\,
    \frac{1}{\Lambda_{\sigma}^{(d)}(q)}
    \right]^{-1}.
    \label{eq. vst exact}
\end{equation}

In the stationary regime, the Lagrange multiplier $v_{\rm st}$ becomes time-independent and is determined by a self-consistency condition analogous to the spherical constraint in the spherical model at equilibrium, corresponding to the $n\to\infty$ limit of the $O(n)$ model. As a result, the stationary two-point correlation function is formally identical to the critical equilibrium correlation function of the long-range spherical model~\cite{cannas2001dynamics}, provided that the microscopic interaction kernels coincide. In particular, comparison of the corresponding normalization conditions leads to the identification $v_{\rm st}=2T_c$, where $T_c$ denotes the critical temperature of the spherical model~\cite{baumann2007kinetics,henkel2011non}.


The explicit value of $v_{\rm st}$ depends on the behavior of the kernel over the whole Brillouin zone and is therefore sensitive to microscopic details. Its existence in the thermodynamic limit, instead, is controlled exclusively by the infrared sector. Using Eq.~(\ref{eq. unified infrared Lambda}),
\begin{equation}
    \int\frac{d^dq}{\Lambda_{\sigma}^{(d)}(q)}
    \sim
    \int_0 dq\,q^{d-1-\beta}\,.
    \label{eq. unified stationary integral}
\end{equation}
The integral converges at the lower limit if and only if
\begin{equation}
    d>\beta\,.
    \label{eq. stationary criterion}
\end{equation}
Thus, a nontrivial stationary state exists in the thermodynamic limit precisely when the effective long-wavelength propagator is transient. 
In terms of the original interaction exponent,
\begin{equation}
    d>\beta
    \quad\Longleftrightarrow\quad
    \begin{cases}
        d>\sigma\,, & 0<\sigma<2\,,\\[1mm]
        d>2\,, & \sigma>2\,.
    \end{cases}
    \label{eq. stationary criterion explicit}
\end{equation}
At $d=\beta$ the divergence is logarithmic, whereas for $d<\beta$ it is algebraic. In both cases $v_{\rm st}\rightarrow0$ in the thermodynamic limit and no nontrivial stationary state exists. When $\sigma>2$, the critical dimension is $d=2$, consistently with the ordinary voter model~\cite{KrapivskyRednerBenNaim2010,HenkelStoimenov2026}.

When $d>\beta$, the large-distance behavior of the stationary correlation function follows from the same infrared sector,
\begin{equation}
    C^F_{\rm st}(q)
    \simeq
    \frac{v_{\rm st}}
    {\mathcal{D}_{\beta}^{(d)}\,|q|^\beta}\,.
    \label{eq. unified stationary structure factor}
\end{equation}
Using the $d$-dimensional Fourier transform of the  Riesz kernel~\cite{stein1970singular},
\begin{equation}
\begin{aligned}
&\int\frac{d^dq}{(2\pi)^d}\,
\frac{e^{i\boldsymbol q\cdot\boldsymbol r}}{|q|^\beta}
=\mathcal R_{d,\beta}\,r^{\beta-d},\\
&\mathcal R_{d,\beta}\equiv
\frac{\Gamma\bigl(\frac{d-\beta}{2}\bigr)}
{2^\beta\pi^{d/2}\Gamma\bigl(\frac{\beta}{2}\bigr)},
\quad 0<\beta<d\,.
\end{aligned}
    \label{eq. Riesz beta}
\end{equation}
we obtain
\begin{equation}
    C_{\rm st}(r)
    \simeq
    \frac{v_{\rm st}}{\mathcal{D}_{\beta}^{(d)}}\,
    \mathcal{R}_{d,\beta}\,
    r^{-(d-\beta)}\,,
    \qquad r\gg1\,.
    \label{eq. unified stationary correlation}
\end{equation}
The stationary correlations decay algebraically, with a universal exponent $d-\beta$ and a kernel-dependent amplitude. In particular, Eq.~\eqref{eq. unified stationary correlation} gives $C_{\rm st}(r)\sim r^{-(d-\sigma)}$ for $0<\sigma<2$ and $C_{\rm st}(r)\sim r^{-(d-2)}$ for $\sigma>2$, and is confirmed by numerically solving Eq.~\eqref{eq. exact discrete} for a two-dimensional model with $\sigma=1.5$ (Fig.~\ref{Fig. steady-state}). These results extend previous findings~\cite{CorberiCastellano2024,CorberiSmaldone2024,CorberiDelloRussoSmaldone2024b,GarofaloLippielloRippa2025}. 

For $2<\sigma<4$, the nonanalytic sector also yields a subleading stationary correction,
\begin{equation}
    C_{\rm st}^{\rm LR}(r)\propto r^{-(d+\sigma-4)}\,.
    \label{eq. stationary long range correction}
\end{equation}

Its derivation, including the treatment of analytic subtractions and marginal even values of $\sigma$, is given in Appendix~\ref{app:stationary correction}. 

The algebraic decay of the stationary correlation function (Eq.~(\ref{eq. unified stationary correlation})) implies the absence of a finite correlation length. In this sense, the stationary state exhibits a critical-like behavior, analogous to that of an equilibrium system at its critical temperature. Although voter dynamics is not governed by a thermodynamic temperature, its approach to stationarity can therefore be compared to the relaxation following a quench to a critical point.


\section{Approach to the stationary state}
\label{sec:approach stationary}

We now discuss the dynamics leading to the stationary states discussed in Sec.~\ref{sec:stationary}, restricting to $d>\beta$.

Starting from Eq.~(\ref{eq. formal Fourier space solution}), it is convenient to introduce the propagator
\begin{equation}
    G(\boldsymbol r,t)
    =
    \int_{\rm BZ}\frac{d^dq}{(2\pi)^d}\,
    e^{i\boldsymbol q\cdot\boldsymbol r}
    e^{-\Lambda_{\sigma}^{(d)}(q)t}\,,
    \label{eq. propagator}
\end{equation}
and its return probability
\begin{equation}
    g(t)\equiv G(0,t)\,.
    \label{eq. return probability}
\end{equation}
The real-space correlation function can then be written as
\begin{equation}
    C(r,t)
    =
    G(r,t)+
    \int_0^t ds\,v(s)\,G(r,t-s)\,,
    \label{eq. correlation in terms of propagator}
\end{equation}
while the constraint in Eq.~\eqref{eq. lagrange constraint} yields the exact relation
\begin{equation}
    1
    =
    g(t)+
    \int_0^t ds\,v(s)\,g(t-s)\,.
    \label{eq. lagrange condition}
\end{equation}

\subsection{Lagrange multiplier in the stationary sector}
\label{subsec:lagrange stationary beta}

At long times the return probability is controlled by Eq.~(\ref{eq. unified infrared Lambda}). With $H_{d-1}=2\pi^{d/2}/\Gamma(d/2)$ denoting the area of the unit hypersphere $\mathcal{H}^{d-1}$,
\begin{equation}
\begin{aligned}
g(t)&\simeq a_{d,\beta}\,t^{-d/\beta},\\
a_{d,\beta}&=\frac{H_{d-1}}{(2\pi)^d}\,
\frac{\Gamma(d/\beta)}
{\beta[\mathcal D_\beta^{(d)}]^{d/\beta}}\,.
\end{aligned}
    \label{eq. return beta}
\end{equation}
For $d>\beta$ this function is integrable. Writing $v(t)=v_{\rm st}+\delta v(t)$ in the exact constraint (Eq.~\eqref{eq. lagrange condition}), the tail $\int_t^\infty g(u)du\sim t^{1-d/\beta}$ gives
\begin{equation}
    v(t)=v_{\rm st}+\mathcal O\!\left(t^{-(d/\beta-1)}\right).
    \label{eq. v approach beta}
\end{equation}
Thus $v(t)$ can be replaced asymptotically by $v_{\rm st}$ in the stationary sector. The initial-condition term is then subleading in the scaling region, and Eq.~\eqref{eq. correlation in terms of propagator} reduces to
\begin{equation}
    C(r,t)\simeq v_{\rm st}\int_0^t du\,G_\beta^{(d)}(r,u),
    \label{eq. late times correlation approaching}
\end{equation}
with
\begin{equation}
    G_\beta^{(d)}(r,t)=\int\frac{d^dq}{(2\pi)^d}
    e^{i\boldsymbol q\cdot\boldsymbol r-\mathcal{D}_{\beta}^{(d)}|q|^\beta t}.
    \label{eq. beta propagator}
\end{equation}

\subsection{Scaling form and growing length}
\label{subsec:unified stationarization}

\begin{figure*}[t]
    \centering
    \includegraphics[width=\textwidth]{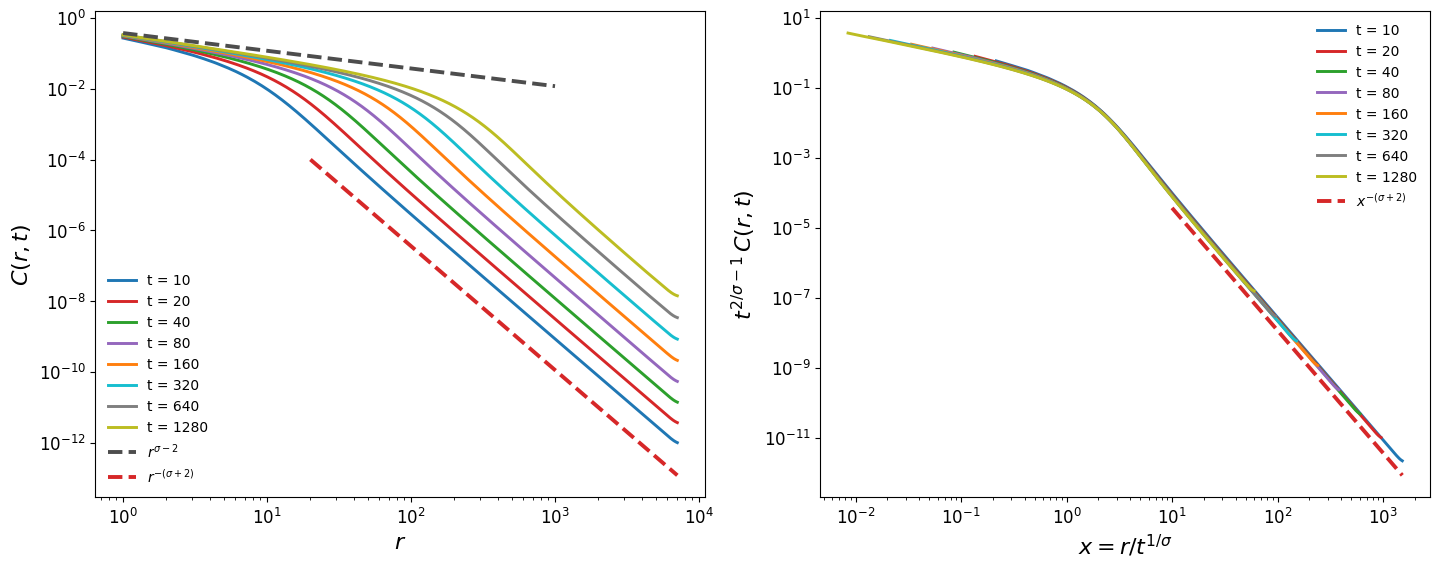}

    \caption{Correlation function $C(r,t)$ for $d=2$
    and $\sigma=1.5$ in a system of size $N=10000^2$.
    Different curves correspond to different times,
    increasing from bottom to top.
    Data are obtained by numerically solving the exact equation in Fourier space,
    Eq.~\eqref{eq. fourier evolution}, with the full kernel in Eq.~(\ref{eq. Lambda}).
    The left panel shows the evolution of the
    correlation function before stationarity,
    together with the predicted algebraic behaviors: the black dotted curve is the expected behavior of the stationary correlation (Eq.~(\ref{eq. unified stationary correlation})), while the red dotted curve is its tail behavior (Eq.~(\ref{eq. tail correlation})).
    The right panel illustrates the scaling collapse
    predicted by Eq.~(\ref{eq. correlation approach st beta}),
    with $x=r/t^{1/\sigma}$.}

    \label{Fig. steady-state}
\end{figure*}
By rescaling
\begin{equation}
    \boldsymbol k
    =
    \left[\mathcal{D}_{\beta}^{(d)}t\right]^{1/\beta}
    \boldsymbol q\,,
\end{equation}
the propagator becomes
\begin{equation}
\begin{aligned}
G_\beta^{(d)}(r,t)
={}&[\mathcal D_\beta^{(d)}t]^{-d/\beta}\\
&\times\mathcal L_{d,\beta}\!\left(
\frac{r}{[\mathcal D_\beta^{(d)}t]^{1/\beta}}\right).
\end{aligned}
    \label{eq. propagator scaling beta}
\end{equation}
where
\begin{equation}
    \mathcal{L}_{d,\beta}(x)
    =
    \int\frac{d^dk}{(2\pi)^d}\,
    e^{i\boldsymbol k\cdot\boldsymbol x-|k|^\beta}\,
    \label{eq. stable density beta}
\end{equation}
is the isotropic $d$-dimensional $\beta$-stable L\'evy density~\cite{metzler2000random}. For $\beta=2$,
\begin{equation}
    \mathcal{L}_{d,2}(x)
    =
    (4\pi)^{-d/2}e^{-x^2/4}\,.
    \label{eq. gaussian stable density}
\end{equation}

Substituting Eq.~(\ref{eq. propagator scaling beta}) into Eq.~(\ref{eq. late times correlation approaching}) and setting $y=u/t$, one obtains
\begin{equation}
\begin{aligned}
C(r,t)\simeq{}&v_{\rm st}
[\mathcal D_\beta^{(d)}]^{-d/\beta}
 t^{1-d/\beta}\\
&\times F_{d,\beta}\!\left(
\frac{r}{[\mathcal D_\beta^{(d)}t]^{1/\beta}}
\right).
\end{aligned}
    \label{eq. correlation approach st beta}
\end{equation}
with
\begin{equation}
    F_{d,\beta}(x)
    =
    \int_0^1dy\,
    y^{-d/\beta}
    \mathcal{L}_{d,\beta}
    \left(xy^{-1/\beta}\right)\,.
    \label{eq. F beta}
\end{equation}
This scaling behavior is confirmed by numerically solving Eq.~\eqref{eq. exact discrete} for a two-dimensional model, with $\sigma=1.5$ (Fig.~\ref{Fig. steady-state}, right panel). The scaling form obtained in Eq.~(\ref{eq. correlation approach st beta}) naturally generalizes the scaling ansatz introduced in Ref.~\cite{HenkelStoimenov2026} for the $d$-dimensional short-range voter model,
\begin{equation}
    C(r,t)=t^{-b}\,F\left(\frac{r}{t^{1/2}}\right)\,,
    \label{eq. henkel}
\end{equation}
For the short-range case, corresponding to $\beta=2$ in our notation, Eq.~(\ref{eq. correlation approach st beta}) gives
\begin{equation}
    b=\frac{d}{2}-1.
\end{equation}
This gives $b>0$ for $d>2$, where the system approaches a nontrivial stationary state; at $d=2$ logarithmic corrections arise, while the coarsening regime $d<2$ has $b=0$. Our result extends this structure to purely long-range copying, with the diffusive exponent $2$ replaced by the effective infrared exponent $\beta$.

The critical long-range spherical model exhibits the same
equal-time scaling behavior of Eq.~(\ref{eq. correlation approach st beta})~\cite{baumann2007kinetics,henkel2011non}.
However, for $0<\sigma<2$, the scaling functions differ
when $\sigma<d<2\sigma$, whereas they coincide at leading
order for $d>2\sigma$, provided that the interaction
kernels are identical.

The related critical-coarsening interpretation, including the fractal exponent $D_f=(d+\beta)/2$, is discussed in Appendix~\ref{app:fractal dimension}.

Eq.~(\ref{eq. correlation approach st beta}) identifies the dynamical length
\begin{equation}
    \xi(t)
    =
    \left[\mathcal{D}_{\beta}^{(d)}t\right]^{1/\beta}
    \sim t^{1/\beta}\,.
    \label{eq. xi beta}
\end{equation}
As anticipated, the evolution towards the stationary state  can be viewed as a quench to a critical point and  $\xi(t)$ can thus be interpreted as the growing length characterizing the \emph{critical} relaxation. This length separates spatial scales that have already reached the stationary profile, $r \ll \xi(t)$, from larger separations that are still evolving, $r \gg \xi(t)$. Appendix~\ref{app:stationarization} establishes this interpretation through the ratio $C(r,t)/C_{\mathrm{st}}(r)$ and provides the explicit diffusive form for $\sigma>2$.

\subsection{Far-distance structure}
\label{subsec:stationary far tails}

\begin{figure}[t]
    \centering
    \includegraphics[width=\columnwidth]{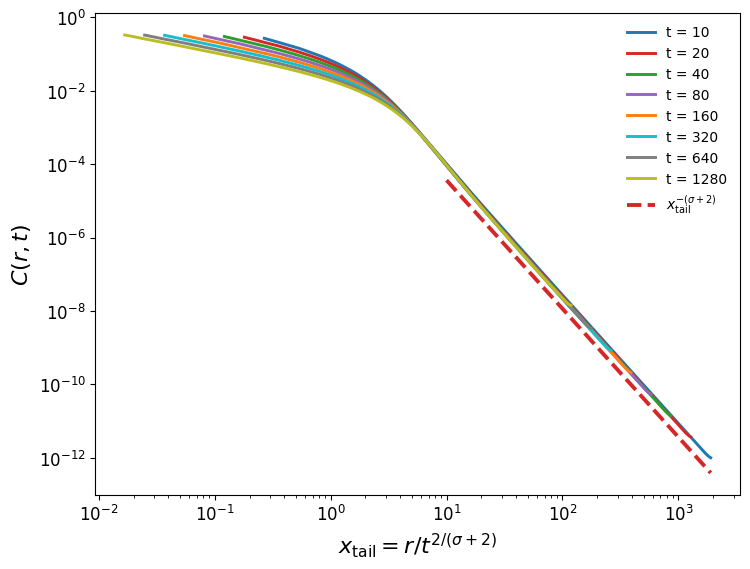}

    \caption{Correlation function $C(r,t)$ for $d=2$
    and $\sigma=1.5$ in a system of size $N=10000^2$.
    Data are obtained by numerically solving the exact equation in Fourier space,
    Eq.~\eqref{eq. fourier evolution}, with the full kernel in Eq.~(\ref{eq. Lambda}).
    The long-range tail of the correlation function obeys the scaling collapse predicted by Eq.~(\ref{eq. tail correlation}),
    with $x_{\rm tail}=r/t^{2/(\sigma+2)}$.}

    \label{Fig. steady-state_tail}
\end{figure}

In both infrared regimes, the algebraic copying kernel produces a far-distance contribution proportional to $t^2r^{-(d+\sigma)}$, but with a different origin. For $0<\sigma<2$, it follows from the large-argument asymptotics of the same L\'evy scaling function $F_{d,\sigma}$ that governs the bulk (Eq.~\eqref{eq. F beta}):
\begin{equation}
    C(r,t)\simeq\frac{v_{\rm st}c_{d,\sigma}D_{\sigma}^{(d)}}{2}
    \frac{t^2}{r^{d+\sigma}},\qquad r\gg\xi(t),
    \label{eq. tail correlation}
\end{equation}
where $c_{d,\sigma}$ is given in Appendix~\ref{app:far distance propagator}. For $\sigma>2$, instead, the bulk propagator is Gaussian and the algebraic tail arises from the subleading nonanalytic $|q|^\sigma$ term in the Fourier generator:
\begin{equation}
    C_{\rm tail}(r,t)\propto t^2r^{-(d+\sigma)},
    \qquad r\gg\xi(t),
    \label{eq. short-range corr tail}
\end{equation}
away from marginal even values of $\sigma$. In the former regime the tail belongs to the leading scaling function; in the latter it is suppressed by $t^{-(\sigma-2)/2}$ at $r\sim\xi(t)$, although it eventually dominates the Gaussian decay at sufficiently large distances. The long-range tail in Eq.~\eqref{eq. tail correlation} is illustrated in Fig.~\ref{Fig. steady-state_tail}.

\section{Non-stationary regime}
\label{sec:nonstationary}

We now consider the coarsening regime $d\leq\beta$, in which the infrared normalization integral in Eq.~(\ref{eq. unified stationary integral}) diverges and no nontrivial stationary state exists. The same infrared exponent that determines the stationary threshold also controls the coarsening dynamics; the nonanalytic $|q|^\sigma$ term must only be restored separately when the far-distance structure is considered for $\sigma>2$.

Let $\widetilde g(p)$ and $\widetilde v(p)$ denote the Laplace transforms of $g(t)$ and $v(t)$. Transforming Eq.~\eqref{eq. lagrange condition} gives the exact relation
\begin{equation}
    \widetilde v(p)=\frac{1}{p\widetilde g(p)}-1.
    \label{eq. v laplace exact}
\end{equation}
Thus the long-time behavior of $v(t)$ follows directly from the small-$p$ return probability.

\subsection{Coarsening below the lower critical dimension:
\texorpdfstring{$d<\beta$}{d < beta}}
\label{subsec:nonstationary below lc}

Set $\varrho=d/\beta<1$. From Eq.~\eqref{eq. return beta}, $g(t)\simeq a_{d,\beta}t^{-\varrho}$, hence $\widetilde g(p)\simeq a_{d,\beta}\Gamma(1-\varrho)p^{\varrho-1}$. Eq.~\eqref{eq. v laplace exact} then yields
\begin{equation}
    v(t)\simeq\kappa_{d,\beta}t^{\varrho-1},\qquad
    \kappa_{d,\beta}=\frac{\sin(\pi\varrho)}{\pi a_{d,\beta}}.
    \label{eq. v recurrent}
\end{equation}
The initial-condition term in Eq.~\eqref{eq. correlation in terms of propagator} is subleading at fixed $r/t^{1/\beta}$. Combining Eq.~\eqref{eq. v recurrent} with the propagator scaling therefore gives
\begin{equation}
    C(r,t)\simeq\mathcal F_{d,\beta}
    \left(\frac{r}{[\mathcal{D}_{\beta}^{(d)}t]^{1/\beta}}\right),
    \qquad d<\beta,
    \label{eq. recurrent scaling}
\end{equation}
with
\begin{equation}
\begin{aligned}
\mathcal F_{d,\beta}(x)
={}&\frac{\sin(\pi\varrho)}
{\pi\mathcal L_{d,\beta}(0)}
\int_0^1dy\,y^{\varrho-1}\\
&\times(1-y)^{-\varrho}\,
\mathcal L_{d,\beta}\!\left(
\frac{x}{(1-y)^{1/\beta}}\right).
\end{aligned}
    \label{eq. recurrent scaling function}
\end{equation}
The coarsening is controlled by the correlation length
\begin{equation}
    L(t)=[\mathcal{D}_{\beta}^{(d)}t]^{1/\beta}\sim t^{1/\beta}.
    \label{eq. recurrent coarsening length}
\end{equation}
Consequently, $L(t)\sim t^{1/\sigma}$ for $0<\sigma<2$ and $L(t)\sim t^{1/2}$ for $\sigma>2$.

\subsubsection{Long-range dominated coarsening,
\texorpdfstring{$0<\sigma<2$}{0 < sigma < 2}}
\label{subsubsec:nonstationary long range}

For $0<\sigma<2$ and $d<\sigma$, the L\'evy tail in Eq.~\eqref{eq. using propagator} implies
\begin{equation}
    C(r,t)\propto t^{1+d/\sigma}r^{-(d+\sigma)},
    \qquad r\gg L(t),
    \label{eq. recurrent long range tail}
\end{equation}
consistent with the one-dimensional results~\cite{CorberiCastellano2024,GarofaloLippielloRippa2025}. Since the temporal amplitude is exactly that implied by $L(t)\sim t^{1/\sigma}$, the bulk and the far tail belong to the same scaling function and no additional growing length is required.

\subsubsection{Diffusive coarsening with long-range copying,
\texorpdfstring{$2<\sigma<4$}{2 < sigma < 4}}
\label{subsubsec:nonstationary short range}

\begin{figure*}[t]
    \centering
    \includegraphics[width=\textwidth]{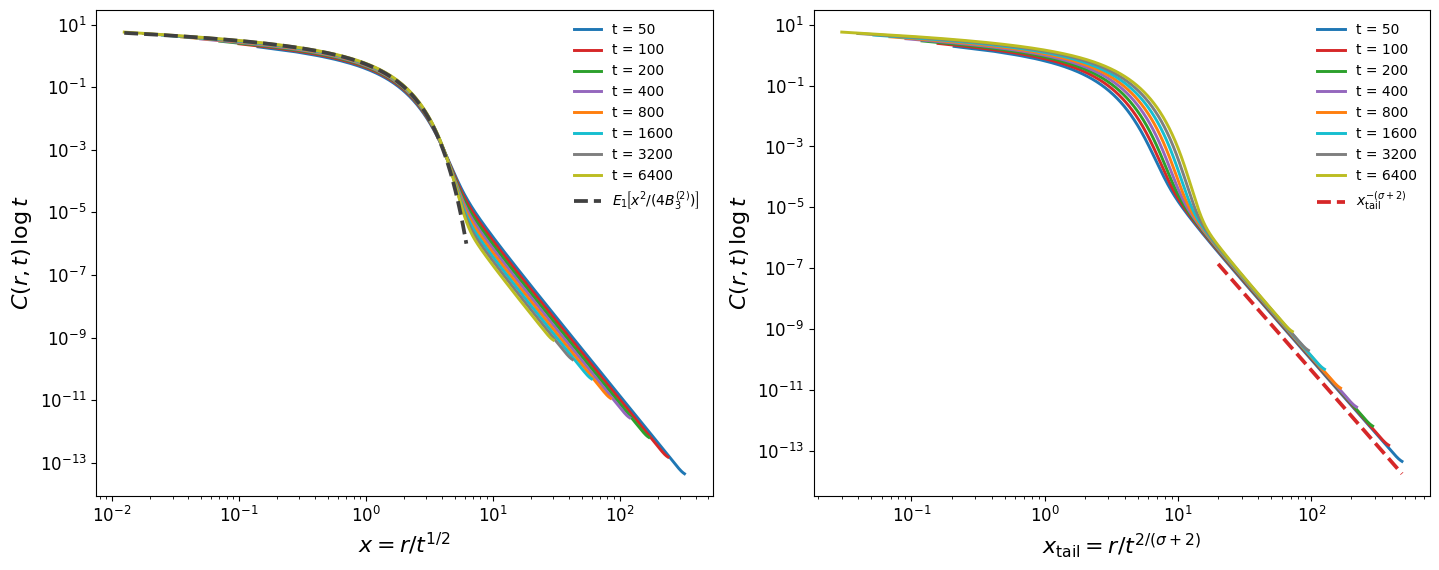}

    \caption{Correlation function for $d=2$ and $\sigma=3$
    in a system of size $N=5000^2$. Different curves correspond to different times in the
    interval $[50,6400]$.
    Data are obtained by numerically solving the exact equation in Fourier space,
    Eq.~\eqref{eq. fourier evolution}, with the full kernel in Eq.~(\ref{eq. Lambda}). The left panel shows data scaled according to the analytical prediction for the bulk diffusive regime, with the dashed line representing the theoretical prediction of Eq.~(\ref{eq. marginal 2d E1}). The right panel displays data scaled according to the long-range tail, with the theoretical behavior given by the power law in Eq.~(\ref{eq. marginal 2d long range tail}).}

    \label{Fig. scaling}
\end{figure*}

For $\sigma>2$ and $d<2$, the leading propagator is Gaussian and Eq.~\eqref{eq. recurrent scaling function} can be evaluated explicitly. With $z=r^2/[4B_{\sigma}^{(d)}t]$,
\begin{equation}
    C_{\rm diff}(r,t)\simeq
    \frac{\Gamma\left(1-\frac d2,z\right)}{\Gamma\left(1-\frac d2\right)},
    \qquad d<2,
    \label{eq. recurrent diffusive scaling}
\end{equation}
which gives the diffusive scaling form of the purely long-range model and agrees, at leading order, with the short-range result~\cite{HenkelStoimenov2026}.

The nonanalytic term does not change the diffusive coarsening length but produces a far-distance correction. Expanding the full propagator around its Gaussian part,
\begin{equation}
    e^{-\Lambda_{\sigma}^{(d)}(q)t}\simeq
    e^{-B_{\sigma}^{(d)}q^2t}
    \left[1-D_{\sigma}^{(d)}t|q|^\sigma+\ldots\right],
    \label{eq. recurrent propagator expansion sr lr}
\end{equation}
and using Eq.~\eqref{eq. v recurrent} gives, away from marginal even $\sigma$,
\begin{equation}
    C_{\rm LR}(r,t)\propto t^{1+d/2}r^{-(d+\sigma)},
    \qquad r\gg L(t).
    \label{eq. recurrent short range tail}
\end{equation}
At $r\sim L(t)$ this term is smaller than the leading scaling function by $t^{-(\sigma-2)/2}$.

The crossover to the algebraic tail occurs beyond the diffusive core. Matching the large-$z$ asymptotics of Eq.~\eqref{eq. recurrent diffusive scaling} with Eq.~\eqref{eq. recurrent short range tail} gives, to leading logarithmic order,
\begin{equation}
    r^*(t)\simeq
    \left[2B_{\sigma}^{(d)}(\sigma-2)t\log t\right]^{1/2},
    \label{eq. recurrent crossover distance}
\end{equation}
with relative corrections of order $\log\log t/\log t$. Hence $r^*(t)/L(t)\sim\sqrt{\log t}$: the long-range tail is asymptotically separated from the diffusive scaling region.

\subsection{Marginal coarsening:
\texorpdfstring{$d=\beta$}{d = beta}}
\label{subsec:nonstationary marginal}

At the critical threshold, $g(t)\sim t^{-1}$ and its Laplace transform is logarithmically singular. Eq.~\eqref{eq. v laplace exact} at late times gives
\begin{equation}
    v(t)\simeq\frac{1}{a_{\beta,\beta}\log t}.
    \label{eq. marginal v}
\end{equation}
The resulting scaling form is
\begin{equation}
    C(r,t)\simeq\frac{1}{\log t}\,
    \mathcal F_{\beta}^{\rm marg}
    \left(\frac{r}{[\mathcal{D}_{\beta}^{(\beta)}t]^{1/\beta}}\right),
    \label{eq. marginal scaling}
\end{equation}
where
\begin{equation}
    \mathcal F_{\beta}^{\rm marg}(x)=
    \frac{1}{\mathcal L_{\beta,\beta}(0)}
    \int_0^1\frac{dy}{y}\,
    \mathcal L_{\beta,\beta}(xy^{-1/\beta}).
    \label{eq. marginal scaling function}
\end{equation}
The growing length remains $L(t)\sim t^{1/\beta}$, while marginality affects the amplitude through the logarithmic factor. This generalizes the behavior previously observed in the short-range voter model at its critical dimension $d_c=2$~\cite{HenkelStoimenov2026}.

For the diffusive marginal case $d=2$, $\sigma>2$, Eq.~\eqref{eq. marginal scaling function} reduces to
\begin{equation}
    C_{\rm diff}(r,t)\simeq\frac{1}{\log t}\,
    E_1\!\left(\frac{r^2}{4B_{\sigma}^{(2)}t}\right),
    \label{eq. marginal 2d E1}
\end{equation}
where $E_1$ is the exponential integral~\cite{CorberiSmaldone2024}. The nonanalytic sector gives
\begin{equation}
    C_{\rm tail}(r,t)\propto\frac{t^2}{\log t}\,r^{-(2+\sigma)},
    \qquad r\gg t^{1/2}.
    \label{eq. marginal 2d long range tail}
\end{equation}
The common $1/\log t$ factor does not alter the matching, so the crossover is again given by Eq.~\eqref{eq. recurrent crossover distance} with $d=2$. 

On the long-range marginal line $d=\sigma<2$,
\begin{equation}
    C(r,t)\simeq\frac{1}{\log t}\,
    \mathcal F_{\sigma}^{\rm marg}
    \left(\frac{r}{[D_{\sigma}^{(\sigma)}t]^{1/\sigma}}\right),
    \label{eq. marginal long range scaling}
\end{equation}
with $L(t)\sim t^{1/\sigma}$. Its far-distance behavior is
\begin{equation}
    C(r,t)\propto\frac{t^2}{\log t}\,r^{-2\sigma},
    \qquad r\gg t^{1/\sigma},
    \label{eq. marginal long range tail}
\end{equation}
and belongs to the same L\'evy scaling sector as the bulk dynamics.

\section{Interface density}

We finally consider the nearest-neighbor interface
density, defined as
\begin{equation}
\begin{aligned}
\rho_{\rm I}(t)
&=\frac{1}{2Nn_1}\sum_i
\sum_{k\in\mathrm{nn}(i)}
\big\langle1-S_i(t)S_k(t)\big\rangle\\
&=\frac{1-C(1,t)}{2}\,,
\end{aligned}
\label{eq:rho_interface}
\end{equation}
where $n_1=2d$.

Its asymptotic behavior can be related directly
to the Lagrange multiplier $v(t)$, whose temporal
evolution has already been determined in the
previous sections.

To establish this relation, we introduce
\begin{equation}
\begin{aligned}
h_1(t)
&\equiv g(t)-G(1,t)\\
&=\int_{\rm BZ}\frac{d^dq}{(2\pi)^d}
[1-\cos(q_1)]\\
&\quad\times
e^{-\Lambda_\sigma^{(d)}(\boldsymbol q)t},
\end{aligned}
\label{eq:interface_kernel}
\end{equation}
where $q_1$ denotes the component of
$\boldsymbol q$ along a nearest-neighbor
lattice direction.

\begin{figure*}[t]
    \centering
    \includegraphics[width=\textwidth]{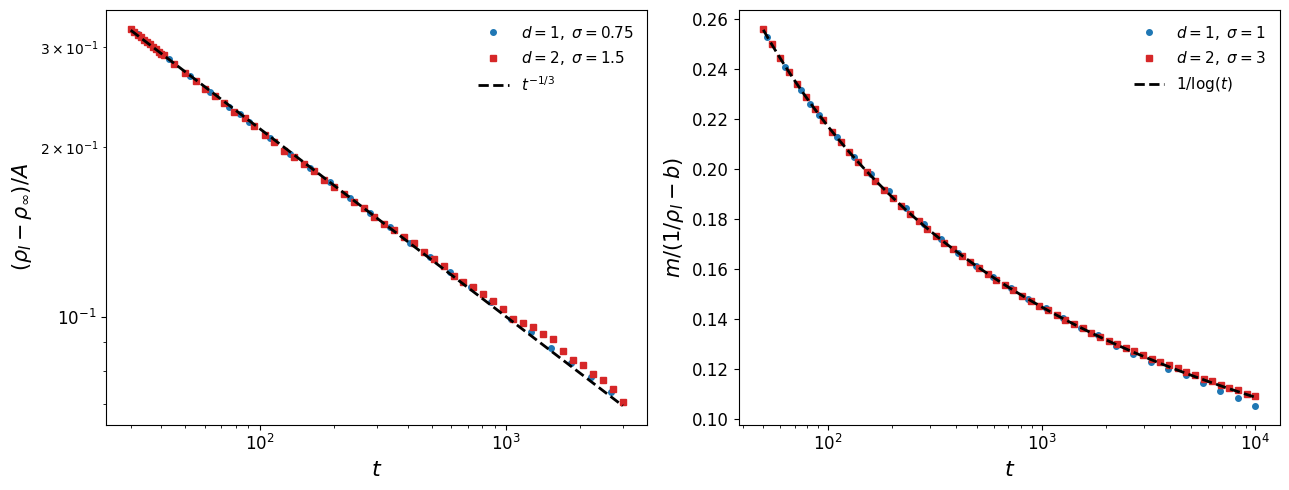}

    \caption{Interface density $\rho_{\rm I}(t)$ as a function of time for systems with effective dimension $d_{\rm eff}=8/3$
(left panel) and $d_{\rm eff}=2$ (right panel). Data are obtained from numerical simulations of one-dimensional systems with $\sigma=0.75$ and two-dimensional systems with $\sigma=1.5$ (left panel), and one-dimensional systems with $\sigma=1$ and two-dimensional systems with $\sigma=3$ (right panel). System sizes are $N=10^6$ in $d=1$ and $N=1000^2$ in $d=2$, and results are averaged over 500 independent realizations.
Data have been appropriately rescaled to eliminate nonuniversal amplitudes, which depend explicitly on the spatial dimension $d$ and the interaction exponent $\sigma$. The rescaled data are consistent with the effective dimension prediction (Eq.~(\ref{eq:rho_interface_deff})).}

    \label{Fig. rho_d_eff}
\end{figure*}

Subtracting Eq.~\eqref{eq. correlation in terms of propagator},
evaluated at $r=1$, from the constraint
in Eq.~\eqref{eq. lagrange condition}
gives the exact relation
\begin{equation}
\begin{aligned}
2\rho_{\rm I}(t)
={}&h_1(t)\\
&+\int_0^t du\,v(t-u)h_1(u).
\end{aligned}
\label{eq:interface_exact_convolution}
\end{equation}

At long times, the infrared behavior of
the generator gives
$h_1(t)\sim t^{-(d+2)/\beta}$.
Since $(d+2)/\beta>1$, the function
$h_1(t)$ is integrable, and we define
\begin{equation}
\begin{aligned}
I_1
&\equiv\int_0^\infty du\,h_1(u)\\
&=\int_{\rm BZ}\frac{d^dq}{(2\pi)^d}
\frac{1-\cos(q_1)}
{\Lambda_\sigma^{(d)}(\boldsymbol q)}.
\end{aligned}
\label{eq:interface_integral}
\end{equation}
The constant $I_1$ is finite and depends
on the microscopic interaction kernel.

For $d\leq\beta$, the function $h_1(t)$
decays faster than $v(t)$, so that the
convolution in
Eq.~\eqref{eq:interface_exact_convolution}
is asymptotically proportional to $v(t)$.
Consequently,
\begin{equation}
\rho_{\rm I}(t)\simeq
\frac{I_1}{2}v(t),
\qquad d\leq\beta.
\label{eq:rho_interface_v}
\end{equation}

For $d>\beta$, instead, $v(t)$ approaches
the stationary value $v_{\rm st}$.
Eq.~\eqref{eq:interface_exact_convolution}
then gives
\begin{equation}
\rho_{\rm I}^{\rm st}
=\frac{I_1}{2}v_{\rm st}.
\label{eq:rho_interface_plateau}
\end{equation}
The same asymptotic argument applies to
the leading corrections to stationarity,
since the contributions associated with
the decay of $h_1(t)$ are subleading.
Thus,
\begin{equation}
\rho_{\rm I}(t)-\rho_{\rm I}^{\rm st}
\simeq\frac{I_1}{2}
[v(t)-v_{\rm st}],
\qquad d>\beta.
\label{eq:rho_interface_relaxation}
\end{equation}

The interface density therefore inherits
the asymptotic temporal behavior of the
Lagrange multiplier, up to a nonuniversal
amplitude. Using the results obtained
in the previous sections, we find
\begin{equation}
\rho_{\rm I}(t)\sim
\begin{cases}
t^{-(1-d/\beta)}, & d<\beta,\\[1mm]
(\log t)^{-1}, & d=\beta,\\[1mm]
\rho_{\rm I}^{\rm st}
+\mathcal O\!\bigl(t^{1-d/\beta}\bigr),
& d>\beta.
\end{cases}
\label{eq:rho_interface_summary}
\end{equation}

\begin{figure}[t]
    \centering
    \includegraphics[width=\columnwidth]{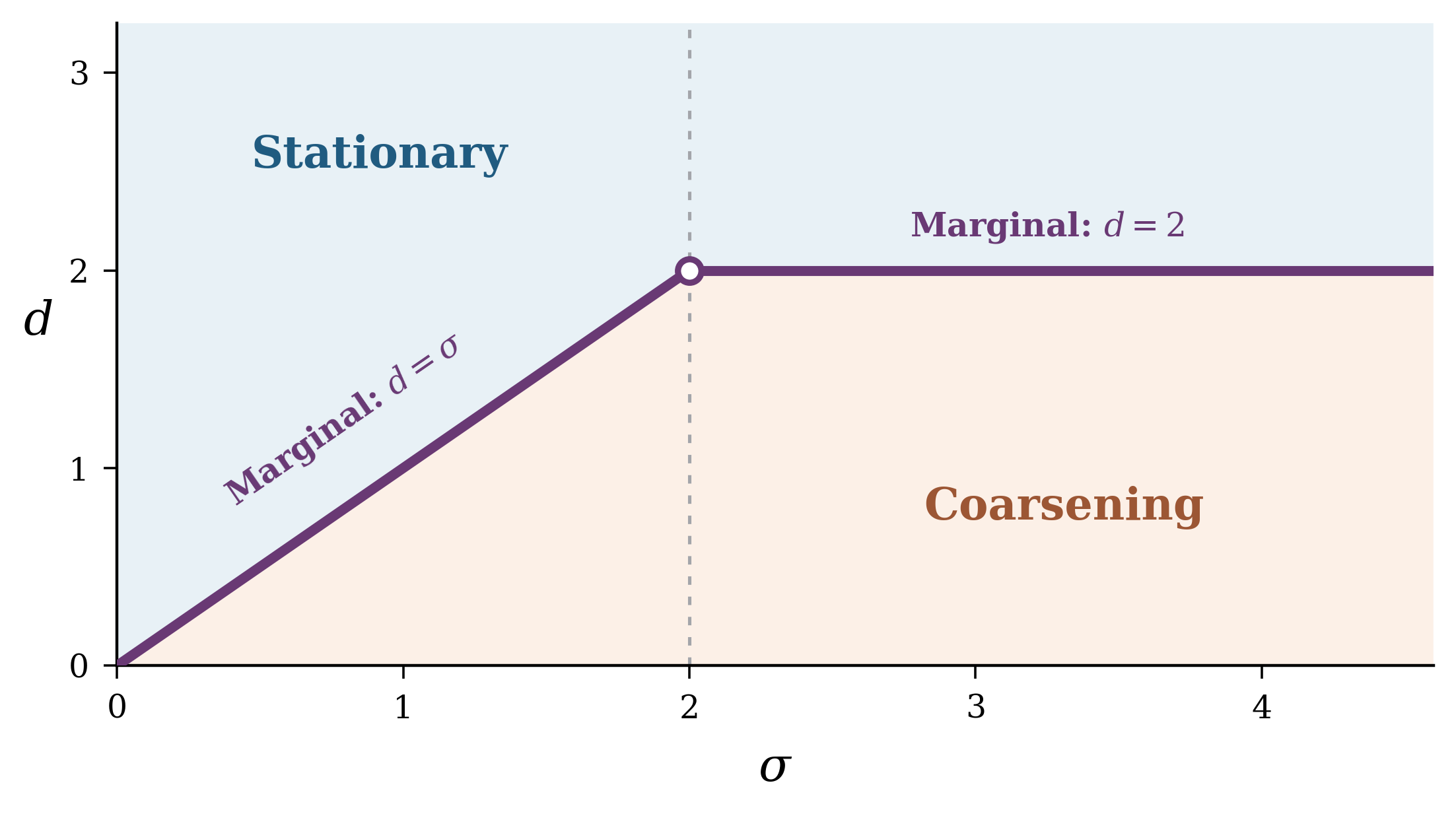}
    \caption{Dynamical regimes of the long-range voter model
    in the $(\sigma,d)$ plane.
    The marginal boundary $d=\beta=\min\{\sigma,2\}$
    separates the coarsening regime ($d<\beta$) from
    the stationary regime ($d>\beta$).
    It consists of two branches: $d=\sigma$ for
    $0<\sigma<2$ and $d=2$ for $\sigma>2$.
    Along this boundary, logarithmic corrections
    characterize the marginal dynamics.
    The special case $\sigma=2$ is not considered.}
    \label{fig:phase_diagram}
\end{figure}

The different dynamical regimes can be summarized
in the $(\sigma,d)$ phase diagram shown in
Fig.~\ref{fig:phase_diagram}.
The boundary separating coarsening from the stationary
regime is given by $d=\beta$, where
$\beta=\min\{\sigma,2\}$.
For $0<\sigma<2$, the marginal line is $d=\sigma$,
whereas for $\sigma>2$ it becomes $d=2$.
Below this boundary ($d<\beta$), the system undergoes
coarsening, while above it ($d>\beta$) a nontrivial
stationary state exists.
On the marginal line, the dynamics exhibits
logarithmic corrections.
The special case $\sigma=2$, which requires a separate
treatment, is excluded from the present analysis.

\section{Effective-dimensional representation}
\label{sec:effective-dimensional}

The results obtained in the previous sections can be recast in terms of an effective spatial dimension. Analogous mappings between long-range systems in dimension $d$ and short-range systems in a generally noninteger dimension have been discussed for equilibrium spin models~\cite{Angelini2014,Defenu2015,Solfanelli2024,pagni2026one}. Here the same construction follows directly from the infrared kernel of the purely long-range voter dynamics.

For $\sigma\neq2$, the dimension $d$ and the infrared exponent $\beta$ enter the leading results primarily through $d/\beta$. We therefore define
\begin{equation}
d_{\rm eff}\equiv\frac{2d}{\beta}
=\max\left\{d,\frac{2d}{\sigma}\right\}.
\label{eq:deff}
\end{equation}
For $\sigma>2$, $d_{\rm eff}=d$, while for $0<\sigma<2$, $d_{\rm eff}=2d/\sigma>d$.

The origin of this mapping is transparent at the level of radial infrared integrals. With $k=q^{\beta/2}$,
\begin{equation}
\int_0 dq\,q^{d-1}f(q^\beta)
=
\frac{2}{\beta}\int_0 dk\,k^{d_{\rm eff}-1}f(k^2),
\label{eq:radial_mapping}
\end{equation}
so that $d/\beta=d_{\rm eff}/2$. Consequently, $d<\beta$, $d=\beta$, and $d>\beta$ map respectively onto $d_{\rm eff}<2$, $d_{\rm eff}=2$, and $d_{\rm eff}>2$. The critical threshold is therefore mapped onto the marginal dimension of the short-range voter model~\cite{KrapivskyRednerBenNaim2010,HenkelStoimenov2026}.

\begin{table*}[t]
\centering
\caption{Comparison between the scaling behavior of the short-range voter model
in dimension $D$, the purely long-range model in physical dimension $d$, and its
effective-dimensional representation. In the last column,
$d_{\rm eff}=2d/\beta$, $\mathcal{R}=r^{\beta/2}$, and
$\mathcal{L}(t)=\ell_\beta(t)^{\beta/2}$. The hatted scaling functions are
defined through the change of scaling variable and are not, in general,
identical to the short-range scaling functions.}
\label{tab:effective_dimension}

\begingroup
\renewcommand{\arraystretch}{1.24}
\setlength{\tabcolsep}{4pt}
\resizebox{\textwidth}{!}{%
\begin{tabular}{llll}
\toprule
Quantity
&
Short-range model in $D$
&
Long-range model in $(d,\beta)$
&
Effective-dimensional form
\\
\midrule

Critical condition
&
$D=2$
&
$d=\beta$
&
$d_{\rm eff}=2$
\\[1.5ex]

Characteristic length
&
$\ell_{\rm SR}(t)\sim t^{1/2}$
&
$\ell_\beta(t)\sim t^{1/\beta}$
&
$\mathcal{L}(t)\sim t^{1/2}$
\\[1.5ex]

Coarsening
&
$\displaystyle
\begin{gathered}
C_{\rm SR}(R,t)
\simeq
F_D^{\rm SR}
\left(\frac{R}{t^{1/2}}\right),
\quad D<2,
\\[1mm]
\rho_{\rm I}^{\rm SR}(t)
\sim
t^{-(1-D/2)}
\end{gathered}$
&
$\displaystyle
\begin{gathered}
C(r,t)
\simeq
F_{d,\beta}
\left(\frac{r}{t^{1/\beta}}\right),
\quad d<\beta,
\\[1mm]
\rho_{\rm I}(t)
\sim
t^{-(1-d/\beta)}
\end{gathered}$
&
$\displaystyle
\begin{gathered}
C(\mathcal{R},t)
\simeq
\widehat F_{d_{\rm eff},\beta}
\left(\frac{\mathcal{R}}{t^{1/2}}\right),
\quad d_{\rm eff}<2,
\\[1mm]
\rho_{\rm I}(t)
\sim
t^{-(1-d_{\rm eff}/2)}
\end{gathered}$
\\[4ex]

Marginal case
&
$\displaystyle
\begin{gathered}
C_{\rm SR}(R,t)
\simeq
\frac{1}{\log t}
F_2^{\rm marg}
\left(\frac{R}{t^{1/2}}\right),
\\[1mm]
\rho_{\rm I}^{\rm SR}(t)
\sim
(\log t)^{-1}
\end{gathered}$
&
$\displaystyle
\begin{gathered}
C(r,t)
\simeq
\frac{1}{\log t}
F_\beta^{\rm marg}
\left(\frac{r}{t^{1/\beta}}\right),
\\[1mm]
\rho_{\rm I}(t)
\sim
(\log t)^{-1}
\end{gathered}$
&
$\displaystyle
\begin{gathered}
C(\mathcal{R},t)
\simeq
\frac{1}{\log t}
\widehat F_\beta^{\rm marg}
\left(\frac{\mathcal{R}}{t^{1/2}}\right),
\\[1mm]
\rho_{\rm I}(t)
\sim
(\log t)^{-1}
\end{gathered}$
\\[4ex]

Approaching stationary
&
$\displaystyle
\begin{gathered}
C_{\rm SR}(R,t)
\sim
t^{1-D/2}
F_D^{\rm SR}
\left(\frac{R}{t^{1/2}}\right),
\quad D>2,
\\[1mm]
\rho_{\rm I}^{\rm SR}(t)
=
\rho_{\rm I,st}^{\rm SR}
+\mathcal O\left(t^{-(D/2-1)}\right)
\end{gathered}$
&
$\displaystyle
\begin{gathered}
C(r,t)
\sim
t^{1-d/\beta}
F_{d,\beta}
\left(\frac{r}{t^{1/\beta}}\right),
\quad d>\beta,
\\[1mm]
\rho_{\rm I}(t)
=
\rho_{\rm I}^{\rm st}
+\mathcal O\left(t^{-(d/\beta-1)}\right)
\end{gathered}$
&
$\displaystyle
\begin{gathered}
C(\mathcal{R},t)
\sim
t^{1-d_{\rm eff}/2}
\widehat F_{d_{\rm eff},\beta}
\left(\frac{\mathcal{R}}{t^{1/2}}\right),
\quad d_{\rm eff}>2,
\\[1mm]
\rho_{\rm I}(t)
=
\rho_{\rm I}^{\rm st}
+\mathcal O\left(t^{-(d_{\rm eff}/2-1)}\right)
\end{gathered}$
\\[4ex]

Stationary correlation
&
$\displaystyle
C_{\rm st}^{\rm SR}(R)
\sim R^{-(D-2)}$
&
$\displaystyle
C_{\rm st}(r)
\sim r^{-(d-\beta)}$
&
$\displaystyle
C_{\rm st}(\mathcal{R})
\sim
\mathcal{R}^{-(d_{\rm eff}-2)}$
\\

\bottomrule
\end{tabular}%
}
\endgroup

\end{table*}

For the spatial scaling it is useful to introduce
\begin{equation}
\mathcal R\equiv r^{\beta/2},\qquad
\mathcal L(t)\equiv\ell_\beta(t)^{\beta/2}\sim t^{1/2},
\label{eq:R_eff}
\end{equation}
where $\ell_\beta$ denotes either the coarsening length $L(t)$ or the growing length $\xi(t)$. The coordinate $\mathcal R$ is defined such that the volume enclosed within a radius $\mathcal R$ scales as $\mathcal R^{d_{\rm eff}}$. The associated scaling variable is $X=\mathcal R/\mathcal L(t)$, and we write $\widehat F_{d_{\rm eff},\beta}(X)\equiv F_{d,\beta}(X^{2/\beta})$. The explicit $\beta$ dependence is retained because the mapping concerns the infrared exponents and classification, not an identity between the complete long-range and short-range scaling functions.

The same representation applies to the interface density. Since $d/\beta=d_{\rm eff}/2$, Eq.~\eqref{eq:rho_interface_summary} becomes
\begin{equation}
\rho_{\rm I}(t)\sim
\begin{cases}
t^{-(1-d_{\rm eff}/2)}, & d_{\rm eff}<2,\\[1mm]
(\log t)^{-1}, & d_{\rm eff}=2,\\[1mm]
\rho_{\rm I}^{\rm st}
+\mathcal O\!\bigl(t^{1-d_{\rm eff}/2}\bigr),
& d_{\rm eff}>2.
\end{cases}
    \label{eq:rho_interface_deff}
\end{equation}
Thus the effective dimension maps the vanishing interface density below the critical threshold, the logarithmic marginal decay, and the approach to a finite stationary plateau onto the corresponding short-range structure~\cite{KrapivskyRednerBenNaim2010,frachebourg1996exact}.


The comparison in Table~\ref{tab:effective_dimension} makes the main consequence of the mapping explicit: the critical threshold, characteristic growth law, temporal scaling exponent, stationary spatial exponent, and interface-density relaxation all acquire the short-range voter form after $d\to d_{\rm eff}$ and $r\to\mathcal R$. In particular, the return probability in Eq.~(\ref{eq. return beta}) becomes $g(t)\sim t^{-d_{\rm eff}/2}$, while the stationary exponent in Eq.~(\ref{eq. unified stationary correlation}) becomes $C_{\rm st}(\mathcal R)\sim\mathcal R^{-(d_{\rm eff}-2)}$. The same behavior is observed in the large-distance critical equilibrium correlator for discrete systems quenched to the critical temperature~\cite{henkel2025physical}.

\section{Discussion and outlook}

In this work we have obtained an exact analytical description of the long-range voter model in arbitrary spatial dimension, showing how its ordering dynamics is governed by the interplay between the spatial dimension \(d\) and the infrared exponent \(\beta=\min\left\{2,\sigma\right\}\). For \(d<\beta\), the system undergoes coarsening, while \(d=\beta\) identifies the marginal regime, characterized by logarithmic corrections. For \(d>\beta\), instead, the dynamics approaches a nontrivial stationary state with algebraically decaying spatial correlations. Our analysis provides a unified description of these regimes, including the characteristic growing lengths and the relaxation of the interface density. We have also shown that the algebraic nature of the copying kernel leaves a signature in the far-distance correlations even when the leading large-scale dynamics is diffusive.

A particularly significant outcome of our analysis is the effective-dimensional correspondence developed in Sec.~\ref{sec:effective-dimensional}. As summarized in Table~\ref{tab:effective_dimension}, the effective dimension introduced in Eq.~(\ref{eq:deff}) allows the main scaling properties of the long-range model to be expressed in terms of those of the short-range voter model. This correspondence encompasses not only the distinction between coarsening, marginal, and stationary regimes, but also the scaling exponents governing spatial correlations, characteristic lengths, and interface-density relaxation. This is an analytical realization of the general framework proposed for the long-range Ising model~\cite{Angelini2014,Defenu2015,Solfanelli2024,pagni2026one}. Moreover, our results also provide an analytical foundation for the effective multidimensional crossover scenario proposed for the one-dimensional long-range voter model~\cite{Rodriguez2011}, while extending it to arbitrary spatial dimension. The correspondence is exact at the level of the infrared scaling exponents, although the associated scaling functions are not, in general, identical.

The analytical framework developed here also opens the way to further investigations. In particular, the approach based on the exact evolution equations and their Fourier-space solution can be naturally extended to two-time correlation and response functions, providing a route to the study of aging properties in arbitrary spatial dimension. This would allow us to develop and generalize the results obtained for the one-dimensional long-range voter model using analytical arguments based on dynamical scaling~\cite{CorberiSmaldone2024Aging}. It would also be interesting to investigate whether the effective-dimensional correspondence established here for equal-time quantities extends to aging exponents and two-time scaling functions. The relevance of these questions is also reflected in the recent interest in aging phenomena in voter dynamics~\cite{Henkel2026Fractal,CorberiSmaldone2024Response}.

\appendix

\numberwithin{equation}{section}
\renewcommand{\theequation}{\thesection\arabic{equation}}

\section{Continuum kernel and its small-momentum expansion}
\label{app:kernel}

We start from the exact evolution equation on the lattice Eq.~(\ref{eq. exact discrete}).
We consider the interaction term
\begin{equation}
\mathcal I(\boldsymbol r,t)
=\sum_p P(l_p)\sum_{k=1}^{n_p}
C(\boldsymbol r-\boldsymbol\ell_{p,k},t),
\label{eq:app_interaction}
\end{equation}
where $\boldsymbol\ell_{p,k}$ is the displacement from a reference
site $i$ to the $k$-th site in its $p$-th proximity shell, and
$|\boldsymbol\ell_{p,k}|=l_p$. 

\begin{figure}[t]
\centering
\includegraphics[width=0.75\linewidth]{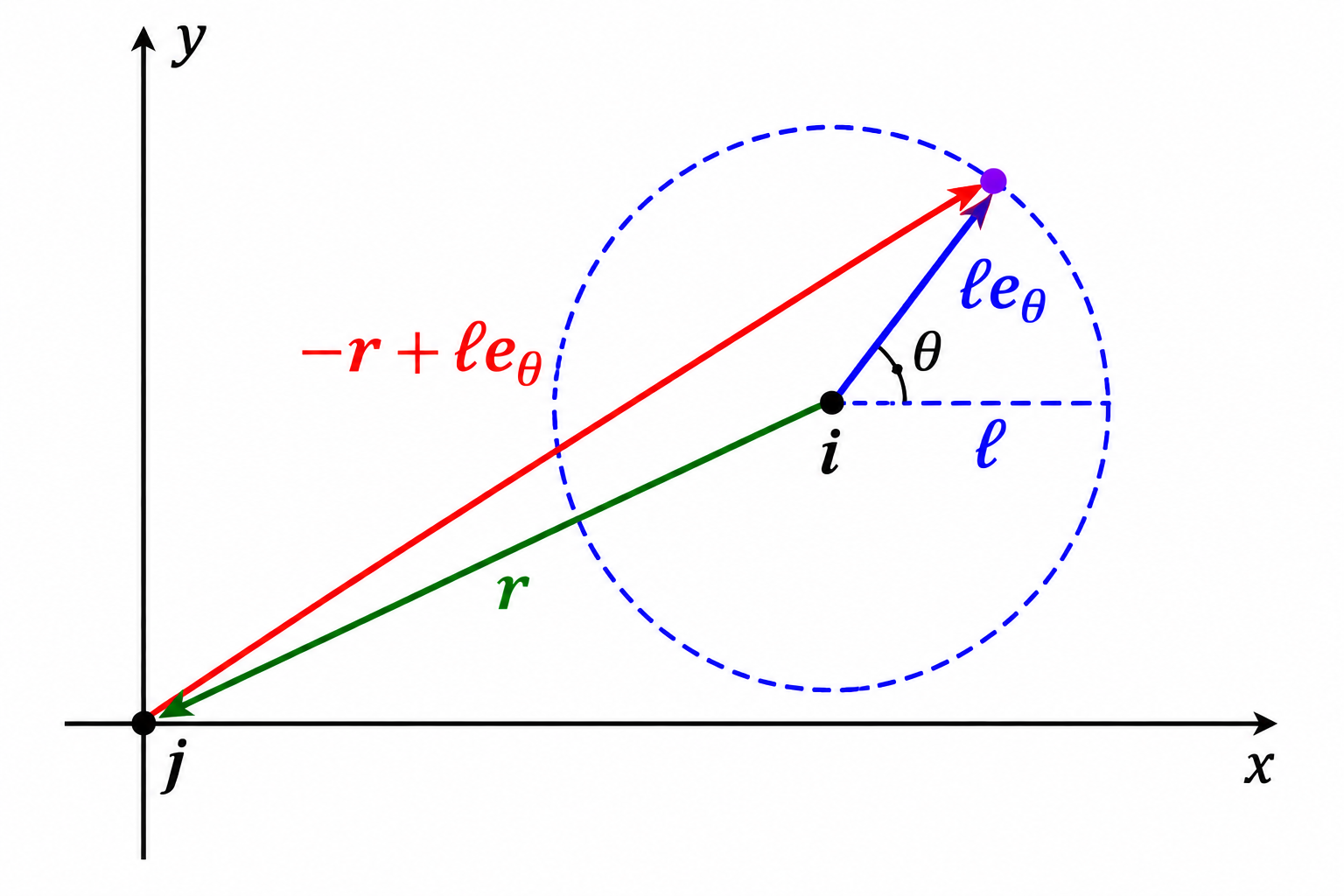}
\caption{Geometrical interpretation of the interaction term in two
 dimensions. Taking site $j$ as the origin, site $i$ is at
 $-\boldsymbol r$. The dashed circle of radius $l$ is centered at
 $i$. The displacement $l\hat{\boldsymbol e}_\theta$ identifies a
 point on this circle, while $-\boldsymbol r+l\hat{\boldsymbol e}_\theta$
 connects the origin to that point. On the lattice only discrete
 points on the circle are allowed.}
\label{fig:geometry}
\end{figure}

The geometrical meaning of Eq.~\eqref{eq:app_interaction} is easily
illustrated in two dimensions. Taking site $j$ as the origin, the
vector $\boldsymbol r=\boldsymbol r_j-\boldsymbol r_i$ places site $i$
at $-\boldsymbol r$. A site at displacement $\boldsymbol\ell$ from
$i$ is then reached from $j$ by the vector
$-\boldsymbol r+\boldsymbol\ell$. Since equal-time correlations are
even, $C(-\boldsymbol r+\boldsymbol\ell,t)
=C(\boldsymbol r-\boldsymbol\ell,t)$. For a given distance $l_p$,
these sites form a discrete shell around $i$. In a continuum
description this shell becomes a circle of radius $l_p$, as shown
in Fig.~\ref{fig:geometry}. The same construction generalizes to
$d$ dimensions, where the circle is replaced by a $(d-1)$-sphere.
At this stage, however, the allowed orientations remain those of
the lattice.

Using the discrete Fourier convention introduced in the main text,
the transform of the shifted correlation is
\begin{equation}
\begin{aligned}
&\sum_{\boldsymbol r}
 e^{-i\boldsymbol q\cdot\boldsymbol r}
 C(\boldsymbol r-\boldsymbol\ell,t)\\
&\quad=\sum_{\boldsymbol u}
 e^{-i\boldsymbol q\cdot(\boldsymbol u+\boldsymbol\ell)}
 C(\boldsymbol u,t)\\
&\quad=e^{-i\boldsymbol q\cdot\boldsymbol\ell}
 C^F(\boldsymbol q,t),
\end{aligned}
\label{eq:app_shift}
\end{equation}
where $\boldsymbol u=\boldsymbol r-\boldsymbol\ell$.
Consequently,
\begin{equation}
\mathcal I^F(\boldsymbol q,t)
=C^F(\boldsymbol q,t)
\sum_pP(l_p)\sum_{k=1}^{n_p}
 e^{-i\boldsymbol q\cdot\boldsymbol\ell_{p,k}}.
\label{eq:app_interaction_fourier}
\end{equation}
The inversion symmetry of the kernel cancels the imaginary part.
Using $\sum_p n_pP(l_p)=1$, we obtain the exact lattice generator
\begin{equation}
\begin{aligned}
\Lambda_\sigma^{(d)}(\boldsymbol q)
&=2\left[1-\sum_pP(l_p)\sum_{k=1}^{n_p}
\cos(\boldsymbol q\cdot\boldsymbol\ell_{p,k})\right]\\
&=2\sum_pP(l_p)\sum_{k=1}^{n_p}
\left[1-\cos(\boldsymbol q\cdot\boldsymbol\ell_{p,k})\right].
\end{aligned}
\label{eq:app_exact_generator}
\end{equation}
Thus, before taking any limit,
\begin{equation}
    \partial_t C^F(\boldsymbol q,t)=-\Lambda_\sigma^{(d)}\boldsymbol q)C^F(\boldsymbol q,t)+v(t).
\label{eq. exact appendix}
\end{equation}
Although each lattice displacement is weighted only by its length,
the set of allowed displacement vectors is not invariant under
arbitrary rotations. Hence the exact generator generally depends
on the components of $\boldsymbol q$, not only on
$q=|\boldsymbol q|$.

We now determine its long-wavelength behavior directly from
Eq.~\eqref{eq:app_exact_generator}. The limit is taken after the
thermodynamic limit, or on finite lattices in the range
$L^{-1}\ll q\ll1$. For the power-law copying kernel, the
infinite-lattice normalization and generator are
\begin{equation}
Z_d(\sigma)=\sum_{\boldsymbol\ell\in\mathbb Z^d\setminus\{\boldsymbol0\}}
\frac{1}{|\boldsymbol\ell|^{d+\sigma}},
\label{eq:app_Z}
\end{equation}
\begin{equation}
\Lambda_\sigma^{(d)}(\boldsymbol q)
=\frac{2}{Z_d(\sigma)}
\sum_{\boldsymbol\ell\in\mathbb Z^d\setminus\{\boldsymbol0\}}
\frac{1-\cos(\boldsymbol q\cdot\boldsymbol\ell)}
{|\boldsymbol\ell|^{d+\sigma}}.
\label{eq:app_infinite_generator}
\end{equation}
At fixed finite $L$, the generator is an analytic trigonometric
polynomial. The nonanalytic powers discussed below emerge when the
thermodynamic limit is taken first.

\subsection{Long-range dominated regime: \texorpdfstring{$0<\sigma<2$}{0 < sigma < 2}.}
In this range the second moment of the copying distribution
diverges, so a direct quadratic expansion of the cosine is not
permitted. We write $\boldsymbol q=\varepsilon\hat{\boldsymbol q}$,
with $\varepsilon=q$ and $|\hat{\boldsymbol q}|=1$, and introduce
$\boldsymbol x=\varepsilon\boldsymbol\ell$. Equation
\eqref{eq:app_infinite_generator} gives
\begin{equation}
\begin{aligned}
&\frac{\Lambda_\sigma^{(d)}
 (\varepsilon\hat{\boldsymbol q})}{\varepsilon^\sigma}\\
&\quad=\frac{2}{Z_d(\sigma)}\varepsilon^d
\sum_{\boldsymbol\ell\ne\boldsymbol0}
\frac{1-\cos(\hat{\boldsymbol q}\cdot
 \varepsilon\boldsymbol\ell)}
 {|\varepsilon\boldsymbol\ell|^{d+\sigma}}.
\end{aligned}
\label{eq:app_riemann}
\end{equation}
For $0<\sigma<2$, the summand defines an integrable function:
near the origin its angularly integrated behavior is
$x^{1-\sigma}$, while at large $x$ it is bounded by a function
proportional to $x^{-1-\sigma}$. Therefore the rescaled lattice
sum converges to the continuum integral
\begin{equation}
\begin{aligned}
&\lim_{\varepsilon\to0}
\frac{\Lambda_\sigma^{(d)}
 (\varepsilon\hat{\boldsymbol q})}{\varepsilon^\sigma}=\frac{2}{Z_d(\sigma)}
\int_{\mathbb R^d}d^dx\,
\frac{1-\cos(\hat{\boldsymbol q}\cdot\boldsymbol x)}
{|\boldsymbol x|^{d+\sigma}}.
\end{aligned}
\label{eq:app_continuum_limit}
\end{equation}
This is the step at which the continuous angular integration
enters. A rotation of the integration variable leaves both the
measure and the denominator unchanged. The integral therefore
does not depend on the orientation of $\hat{\boldsymbol q}$.

To evaluate it, we choose the polar axis along
$\hat{\boldsymbol q}$, so that
$\hat{\boldsymbol q}\cdot\boldsymbol x=x\cos\theta$. The
angular element in $d\ge2$ dimensions is
\begin{equation}
d\Omega_{d-1}
=(\sin\theta)^{d-2}\,d\theta\,d\Omega_{d-2}.
\end{equation}
The integration over the remaining angles gives the area
$H_{d-2}=2\pi^{(d-1)/2}/\Gamma((d-1)/2)$ of the unit
$(d-2)$-sphere. We introduce the angular transform
\begin{equation}
\begin{aligned}
\mathcal A_d(x)
&\equiv\int_{\mathbb S^{d-1}}d\Omega_{d-1}
 e^{-ix\cos\theta}\\
&=H_{d-2}\int_0^\pi d\theta\,
 (\sin\theta)^{d-2}e^{-ix\cos\theta}.
\end{aligned}
\label{eq:app_angular_integral}
\end{equation}
Setting $u=\cos\theta$, this becomes
\begin{equation}
\mathcal A_d(x)
=H_{d-2}\int_{-1}^{1}du\,
(1-u^2)^{(d-3)/2}e^{-ixu}.
\end{equation}
Using the integral representation
\begin{equation}
\begin{aligned}
&\int_{-1}^{1}du\,(1-u^2)^{\nu-1/2}e^{-ixu}\\
&\qquad=\sqrt{\pi}\,\Gamma(\nu+1/2)
\left(\frac{2}{x}\right)^\nu J_\nu(x),
\end{aligned}
\end{equation}
with $\nu=d/2-1$, we finally obtain
\begin{equation}
\mathcal A_d(x)
=(2\pi)^{d/2}x^{1-d/2}J_{d/2-1}(x).
\label{eq. angular transform}
\end{equation}
Here $J_\nu$ denotes the Bessel function of the first kind.
For $d=2$ this reduces to $\mathcal A_2(x)=2\pi J_0(x)$;
in $d=1$ the same angular transform follows directly by
summing over the two directions. Writing
$H_{d-1}=2\pi^{d/2}/\Gamma(d/2)$ for the area of the
unit $(d-1)$-sphere, its small-$x$ expansion is
\begin{equation}
\mathcal A_d(x)=H_{d-1}\left[
1-\frac{x^2}{2d}+\frac{x^4}{8d(d+2)}+O(x^6)
\right].
\label{eq:app_angular_smallx}
\end{equation}
Equation~\eqref{eq:app_continuum_limit} therefore yields
\begin{equation}
\begin{aligned}
\Lambda_\sigma^{(d)}(\boldsymbol q)
&=D_\sigma^{(d)}q^\sigma+o(q^\sigma),\\
D_\sigma^{(d)}
&=\frac{2}{Z_d(\sigma)}\int_0^\infty
\frac{dx}{x^{1+\sigma}}
\left[H_{d-1}-\mathcal A_d(x)\right],
\end{aligned}
\label{eq:app_D_small_sigma}
\end{equation}
for $0<\sigma<2$. The coefficient is finite and positive. Notice that this
continuum integral gives the exact leading infrared coefficient
of the lattice generator: it is obtained as the limit of the
rescaled lattice sum, not by replacing the microscopic model
with a continuum one at all distances.

\subsection{Finite-second-moment regime: \texorpdfstring{$\sigma>2$}{sigma > 2}.}
When $\sigma>2$, the second moment is finite. Expanding the
cosine inside Eq.~\eqref{eq:app_infinite_generator} gives
\begin{equation}
\Lambda_\sigma^{(d)}(\boldsymbol q)
=\sum_{\mu,\nu=1}^{d}q_\mu q_\nu M_{\mu\nu}
+o(q^2),
\label{eq:app_second_moment_tensor}
\end{equation}
where
\begin{equation}
M_{\mu\nu}=\frac{1}{Z_d(\sigma)}
\sum_{\boldsymbol\ell\ne\boldsymbol0}
\frac{\ell_\mu\ell_\nu}{|\boldsymbol\ell|^{d+\sigma}}.
\end{equation}
Reflections of the hypercubic lattice imply that the
off-diagonal elements vanish. Permutations of the coordinate
axes make all diagonal elements equal. Hence
$M_{\mu\nu}=B_\sigma^{(d)}\delta_{\mu\nu}$, with
\begin{equation}
B_\sigma^{(d)}=
\frac{1}{dZ_d(\sigma)}
\sum_{\boldsymbol\ell\ne\boldsymbol0}
|\boldsymbol\ell|^{2-d-\sigma},
\qquad\sigma>2.
\label{eq:app_B_exact}
\end{equation}
Thus $\Lambda_\sigma^{(d)}(\boldsymbol q)
=B_\sigma^{(d)}q^2+o(q^2)$. Importantly,
Eq.~\eqref{eq:app_B_exact} is the exact coefficient for the
lattice model. 

The algebraic tail of the copying kernel also generates a
nonanalytic correction~\cite{cannas2001dynamics}. For $2<\sigma<4$ we subtract the
quadratic term before taking the continuum limit:
\begin{equation}
\begin{aligned}
&\Lambda_\sigma^{(d)}(\boldsymbol q)
-B_\sigma^{(d)}q^2\\
&\quad=\frac{2}{Z_d(\sigma)}
\sum_{\boldsymbol\ell\ne\boldsymbol0}
\frac{1-\cos(\boldsymbol q\cdot\boldsymbol\ell)
-(\boldsymbol q\cdot\boldsymbol\ell)^2/2}
{|\boldsymbol\ell|^{d+\sigma}}.
\end{aligned}
\label{eq:app_subtracted_sum}
\end{equation}
The rescaling used in Eq.~\eqref{eq:app_riemann} now gives
\begin{equation}
\begin{aligned}
\Lambda_\sigma^{(d)}(\boldsymbol q)
&=B_\sigma^{(d)}q^2
+D_\sigma^{(d)}q^\sigma+o(q^\sigma),\\
D_\sigma^{(d)}
&=\frac{2}{Z_d(\sigma)}\int_0^\infty
\frac{dx}{x^{1+\sigma}}
\bigg[H_{d-1}-\mathcal A_d(x) \\
&\quad-\frac{H_{d-1}}{2d}x^2\bigg],
\end{aligned}
\label{eq:app_D_large_sigma}
\end{equation}
valid for $2<\sigma<4$. The subtraction makes the integral
convergent at the origin; the condition $\sigma>2$ ensures
convergence at infinity. In this range $D_\sigma^{(d)}<0$.

For completeness, when $\sigma>4$ is not an even integer,
higher-order analytic contributions must also be removed
before extracting the nonanalytic correction. Let
$m_\sigma=\lfloor\sigma/2\rfloor$ and define
\begin{equation}
R_{m_\sigma}(u)=1-\cos u
-\sum_{m=1}^{m_\sigma}
\frac{(-1)^{m+1}u^{2m}}{(2m)!}.
\end{equation}
The same argument then gives
\begin{equation}
\Lambda_\sigma^{(d)}(\boldsymbol q)
=\sum_{m=1}^{m_\sigma}Q_{2m}(\boldsymbol q)
+D_\sigma^{(d)}q^\sigma+o(q^\sigma),
\label{eq:app_general_expansion}
\end{equation}
where
\begin{equation}
Q_{2m}(\boldsymbol q)
=\frac{2(-1)^{m+1}}{(2m)!Z_d(\sigma)}
\sum_{\boldsymbol\ell\ne\boldsymbol0}
\frac{(\boldsymbol q\cdot\boldsymbol\ell)^{2m}}
{|\boldsymbol\ell|^{d+\sigma}},
\end{equation}
and
\begin{equation}
D_\sigma^{(d)}
=\frac{2}{Z_d(\sigma)}
\int_{\mathbb R^d}d^dx\,
\frac{R_{m_\sigma}(x_1)}
{|\boldsymbol x|^{d+\sigma}}.
\label{eq:app_D_general}
\end{equation}
The polynomial $Q_2(\boldsymbol q)=B_\sigma^{(d)}q^2$
is isotropic by hypercubic symmetry, whereas
higher-order polynomials such as $Q_4$ need not be.
Therefore, the expression containing only $q^2$ and
$q^\sigma$ identifies the leading analytic and
nonanalytic sectors, but is not a complete expansion
ordered by powers of $q$ when $\sigma>4$.

For non-even $\sigma$, the continuum coefficient
can also be written as
\begin{equation}
D_\sigma^{(d)}
=-\frac{2^{1-\sigma}\pi^{d/2}
\Gamma(-\sigma/2)}
{Z_d(\sigma)\,\Gamma((d+\sigma)/2)}.
\label{eq:app_D_gamma}
\end{equation}

\subsection{Marginal even values}
At even values $\sigma=2n$, with integer $n\geq1$,
the moment of order $2n$ becomes marginally divergent.
This includes $\sigma=2$, where the divergence affects
the leading quadratic term, as well as
$\sigma=4,6,\ldots$, where it appears at subleading
orders.

To determine the resulting logarithmic contribution,
we first consider the moment of order $2n$ with
a radial cutoff $R$:
\begin{equation}
M_{2n}(R)
=
\sum_{0<|\boldsymbol\ell|\leq R}
\frac{|\boldsymbol\ell|^{2n}}
{|\boldsymbol\ell|^{d+2n}}.
\end{equation}
At large $R$, its leading behavior is obtained from
the corresponding continuum integral. Introducing
spherical coordinates gives
\begin{equation}
\begin{aligned}
M_{2n}(R)
&=H_{d-1}\int_1^R dr\,
r^{d-1}\frac{r^{2n}}{r^{d+2n}}
+O(1)\\
&=H_{d-1}\int_1^R\frac{dr}{r}+O(1)\\
&=H_{d-1}\log R+O(1),
\end{aligned}
\end{equation}
where
$H_{d-1}=2\pi^{d/2}/\Gamma(d/2)$
is the area of the unit $(d-1)$-sphere.

The logarithmic divergence is therefore independent
of the spatial dimension: the factor $r^{d-1}$
from the radial measure cancels the dimensional
dependence $r^{-d}$ of the interaction kernel.
The coefficient of the divergence, however, retains
a dependence on $d$.

We now evaluate the corresponding contribution
to the small-momentum expansion of the generator.
At $\sigma=2n$, the analytic terms of order lower
than $2n$ are finite and can be subtracted:
\begin{equation}
\begin{aligned}
&\Lambda_{2n}^{(d)}(\boldsymbol q)
-\sum_{m=1}^{n-1}Q_{2m}(\boldsymbol q)\\
&\qquad =
\frac{2}{Z_d(2n)}
\sum_{\boldsymbol\ell\ne\boldsymbol0}
\frac{R_{n-1}(\boldsymbol q\cdot\boldsymbol\ell)}
{|\boldsymbol\ell|^{d+2n}}.
\end{aligned}
\label{eq:app_even_subtraction}
\end{equation}
Here the sum over $m$ is understood to be empty
when $n=1$. For small arguments,
\begin{equation}
R_{n-1}(u)
=
\frac{(-1)^{n+1}}{(2n)!}u^{2n}
+O(u^{2n+2}).
\end{equation}

We split
the lattice sum into displacements
$|\boldsymbol\ell|\leq R$ and
$|\boldsymbol\ell|>R$, choosing $R\sim q^{-1}$.
The Taylor expansion applies in the first region,
while the remaining contribution is $O(q^{2n})$.
Writing $\boldsymbol q=q\hat{\boldsymbol q}$,
we obtain
\begin{equation}
\begin{aligned}
&\Lambda_{2n}^{(d)}(\boldsymbol q)
-\sum_{m=1}^{n-1}Q_{2m}(\boldsymbol q)\\
&\quad =
\frac{2(-1)^{n+1}q^{2n}}
{(2n)!Z_d(2n)}
\sum_{0<|\boldsymbol\ell|\leq R}
\frac{(\hat{\boldsymbol q}\cdot
\boldsymbol\ell)^{2n}}
{|\boldsymbol\ell|^{d+2n}}
+O(q^{2n}).
\end{aligned}
\end{equation}

The leading divergence of this sum can be evaluated
by separating the radial and angular integrations:
\begin{equation}
\begin{aligned}
&\sum_{0<|\boldsymbol\ell|\leq R}
\frac{(\hat{\boldsymbol q}\cdot
\boldsymbol\ell)^{2n}}
{|\boldsymbol\ell|^{d+2n}}\\
&\quad =
\int_1^R\frac{dr}{r}
\int_{S^{d-1}}d\Omega\,
(\hat{\boldsymbol q}\cdot
\hat{\boldsymbol r})^{2n}
+O(1).
\end{aligned}
\end{equation}

The angular integral is isotropic and can be
evaluated exactly:
\begin{equation}
\begin{aligned}
I_{d,n}
&\equiv
\int_{S^{d-1}}d\Omega\,
(\hat{\boldsymbol q}\cdot
\hat{\boldsymbol r})^{2n}\\
&=
H_{d-1}
\frac{(2n-1)!!}
{d(d+2)\cdots(d+2n-2)}\\
&=
\frac{2\pi^{d/2}(2n)!}
{2^{2n}n!\,\Gamma(d/2+n)}.
\end{aligned}
\label{eq:app_even_angular}
\end{equation}

Using $\log R=\log(1/q)+O(1)$, we finally obtain the general expansion at even interaction exponents:
\begin{equation}
\Lambda_{2n}^{(d)}(\boldsymbol q)
=
\sum_{m=1}^{n-1}Q_{2m}(\boldsymbol q)
+\mathcal K_{d,n}q^{2n}\log(1/q)
+O(q^{2n})
\label{eq:app_even_expansion}
\end{equation}
where
\begin{equation}
\mathcal K_{d,n}
=
\frac{(-1)^{n+1}2^{2-2n}\pi^{d/2}}
{n!\,Z_d(2n)\,\Gamma(d/2+n)}.
\label{eq:app_even_coefficient}
\end{equation}

This expression shows that logarithmic corrections arise at all even values $\sigma=2,4,6,\ldots$. The logarithmically enhanced contribution is isotropic, even when the preceding analytic terms are not.

In particular, for $n=1$, that is, $\sigma=2$, the analytic sum is absent, and Eq.~\eqref{eq:app_even_expansion} reduces to
\begin{equation}
\Lambda_2^{(d)}(q)
=
A_d\, q^2\log(1/q),
\qquad
A_d=\frac{H_{d-1}}{dZ_d(2)}.
\label{eq:app_sigma2}
\end{equation}

For $n\geq2$, the logarithmic correction instead
appears beyond the leading quadratic term and
does not modify the diffusive infrared scaling.

Combining the above results, the small-momentum
expansion of the lattice generator can be summarized as
\begin{equation}
\Lambda_\sigma^{(d)}(q)=
\begin{cases}
D_\sigma^{(d)}q^\sigma+o(q^\sigma),
&0<\sigma<2,\\[4pt]
A_d\,q^2\log(1/q)+O(q^2),
&\sigma=2,\\[4pt]
B_\sigma^{(d)}q^2+D_\sigma^{(d)}q^\sigma
+o(q^\sigma),
&2<\sigma<4,\\[4pt]
B_\sigma^{(d)}q^2+o(q^2),
&\sigma\geq4.
\end{cases}
\label{eq:app_infrared_summary}
\end{equation}
In all three regimes the leading term depends only on
$q=|\boldsymbol q|$. Rotational invariance is therefore
recovered at the level of the infrared generator, even though
it is not an exact symmetry of the lattice at finite
wavevectors.

When $\sigma\neq2$, the ordering of terms in the limit $|q|\rightarrow0$ does not coincide with their role in the large-distance structure of the correlation function. For $0<\sigma<2$, one has $|q|^\sigma\gg q^2$ as $q\rightarrow0$, so that the nonanalytic contribution controls the leading long-wavelength dynamics. Conversely, for $\sigma>2$, $q^2\gg|q|^\sigma$ as $q\rightarrow0$, and the analytic sector determines the leading bulk scaling.

The nonanalytic term nevertheless remains physically relevant in this last case. Analytic powers of $q^2$ correspond in real space to local differential operators, whereas the Fourier transform of a nonanalytic power behaves algebraically,
\begin{equation}
    \mathcal{F}^{-1}\left[|q|^\sigma\right]\propto r^{-d-\sigma}\,,
\end{equation}
away from the origin. Therefore, although $|q|^\sigma$ is subleading in the small-$q$ expansion for $\sigma>2$, it controls the asymptotic algebraic tail of the correlation function at sufficiently large $r$. 

For the fully disordered initial condition,
$C^F(\boldsymbol q,0)=1$, the formal solution to the Fourier-space differential equation (Eq.~(\ref{eq. exact appendix})), 
\begin{equation}
    \begin{aligned}
        C^F(q,t)={}&e^{-\Lambda_{\sigma}^{(d)}(q)t}\\
        &+\int_0^t dt'\,v(t')\,
        e^{-\Lambda_{\sigma}^{(d)}(q)(t-t')}\,,
    \end{aligned}    
\end{equation}
shows that the leading scaling contribution to
$C^F(\boldsymbol q,t)$ inherits this isotropy. The
corresponding real-space scaling function depends on the
separation only through $r=|\boldsymbol r|$, up to lattice
and finite-size corrections. This conclusion concerns the
long-wavelength scaling limit and does not require the
microscopic correlation function to be radial.


\section{Asymptotic long-range propagator}
\label{app:far distance propagator}

For $0<\sigma<2$, the asymptotic tail of the L\'evy propagator is
\begin{equation}
    G_\sigma^{(d)}(r,t)\simeq c_{d,\sigma}\,
    \frac{D_{\sigma}^{(d)}t}{r^{d+\sigma}},
    \qquad r\gg\xi(t),
    \label{eq. using propagator}
\end{equation}
with
\begin{equation}
    c_{d,\sigma}=\frac{2^\sigma\Gamma[(d+\sigma)/2]}
    {\pi^{d/2}|\Gamma(-\sigma/2)|}.
    \label{eq. c dsigma}
\end{equation}
Inserting this expression into Eq.~\eqref{eq. late times correlation approaching} yields Eq.~\eqref{eq. tail correlation}. For $\sigma>2$, the Gaussian bulk contribution is generated by the leading $q^2$ term, whereas the subleading nonanalytic $|q|^\sigma$ contribution produces the algebraic far-distance correction in Eq.~\eqref{eq. short-range corr tail}. Analytic higher-order terms do not generate such a tail away from the origin.

\section{Subleading stationary long-range contribution}
\label{app:stationary correction}
\label{subsec:stationary long range correction}

The unified representation above determines the leading stationary behavior, but for $\sigma>2$ it does not exhaust the spatial structure. In this regime the analytic $q^2$ contribution is dominant, while the nonanalytic $|q|^\sigma$ term generated by the algebraic copying kernel survives as a subleading correction. Retaining both sectors,
\begin{equation}
    C^F_{\rm st}(q)
    =
    \frac{v_{\rm st}}
    {B_{\sigma}^{(d)}\,q^2
    +D_{\sigma}^{(d)}\,|q|^\sigma+\ldots}\,.
    \label{eq. stationary full short range}
\end{equation}
For $2<\sigma<4$, the small-$q$ expansion gives
\begin{equation}
\begin{aligned}
C^F_{\rm st}(q)\simeq{}&
\frac{v_{\rm st}}{B_\sigma^{(d)}}\,q^{-2}\\
&-\frac{v_{\rm st}D_\sigma^{(d)}}
{[B_\sigma^{(d)}]^2}|q|^{\sigma-4}
+\cdots\,.
\end{aligned}
    \label{eq. stationary short range expansion}
\end{equation}
The first term produces the leading short-range-like decay $r^{-(d-2)}$, whereas the nonanalytic contribution gives Eq.~\eqref{eq. stationary long range correction}. Since $\sigma>2$, one has $d+\sigma-4>d-2$, and the long-range contribution decays faster than the leading stationary term. For larger values of $\sigma$, additional analytic even powers of $q$ may occur before the first nonanalytic term. Such analytic terms correspond to local contributions and do not alter the large-distance algebraic correction in Eq.~(\ref{eq. stationary long range correction}). 

At $\sigma=2$ the analytic and nonanalytic sectors merge already at leading order, producing logarithmic corrections.

\section{Critical-coarsening interpretation}
\label{app:fractal dimension}

A closely related interpretation was proposed in Ref.~\cite{CorberiDelloRussoSmaldone2024b} for the three-dimensional long-range voter model, where the scaling behavior of the correlation function was discussed in analogy with critical coarsening~\cite{corberi2004effective}. In this framework, the correlation function is written as
\begin{equation}
    C(r,t)\sim L(t)^{-2(d-D_f)}
    f\left(\frac{r}{L(t)}\right),
\end{equation}
where $D_f$ denotes the fractal dimension of the correlated regions. In our case, using $L(t)\sim t^{1/\beta}$, Eq.~(\ref{eq. correlation approach st beta}) can be rewritten as
\begin{equation}
    C(r,t)\sim L(t)^{-(d-\beta)}
    F\left(\frac{r}{L(t)}\right).
\end{equation}
Comparison with the critical-coarsening form therefore gives
\begin{equation}
    2(d-D_f)=d-\beta\,,
\end{equation}
and hence
\begin{equation}
    D_f=\frac{d+\beta}{2}\,.
    \label{eq. fractal dimension}
\end{equation}
We interpret $D_f$ as the effective fractal exponent associated with this critical-coarsening scaling form. In $d=3$ it gives $(3+\sigma)/2$ for $0<\sigma<2$ and $5/2$ for $\sigma>2$, recovering the results obtained in Ref.~\cite{CorberiDelloRussoSmaldone2024b}.

This critical-coarsening interpretation can also be expressed in terms of the effective-dimension representation discussed in Sec.~\ref{sec:effective-dimensional}. Eq.~(\ref{eq. fractal dimension}), written in terms of $d_{\rm eff}$ and $\mathcal{R}$, gives
\begin{equation}
D_f=\frac{\beta}{4}(d_{\rm eff}+2),\qquad
D_f^{(\mathcal R)}\equiv\frac{2D_f}{\beta}=\frac{d_{\rm eff}+2}{2}.
\label{eq:Df_eff}
\end{equation}
The second expression has the same dimensional dependence as the corresponding short-range result. As outlined in Sec.~\ref{sec:effective-dimensional}, this should be understood as a scaling identification rather than an independent geometrical determination of a fractal set.

\section{Stationarization profile and explicit diffusive limit}
\label{app:stationarization}

Since the final stationary state is algebraically correlated, $\xi(t)$ is not an equilibrium correlation length associated with an exponential cutoff. It is instead the dynamical length separating spatial scales which have already reached their stationary form from those which have not. 

This interpretation follows directly from the ratio between the time-dependent and stationary correlations. Introducing
\begin{equation}
    \tau
    =
    \frac{\mathcal{D}_{\beta}^{(d)}\,t}{r^\beta}\,,
    \label{eq. tau beta}
\end{equation}
Eq.~(\ref{eq. late times correlation approaching}) can be rewritten as
\begin{equation}
    C(r,t)
    =
    \frac{v_{\rm st}}{\mathcal{D}_{\beta}^{(d)}}\,
    r^{\beta-d}\,
    H_{d,\beta}(\tau)\,,
    \label{eq. H representation beta}
\end{equation}
where
\begin{equation}
    H_{d,\beta}(\tau)
    =
    \int_0^\tau dy\,
    y^{-d/\beta}
    \mathcal{L}_{d,\beta}
    \left(y^{-1/\beta}\right)\,.
    \label{eq. H beta}
\end{equation}
For $d>\beta$ the integral converges as $\tau\rightarrow\infty$, and
\begin{equation}
    H_{d,\beta}(\infty)
    =
    \mathcal{R}_{d,\beta}\,,
    \label{eq. H infinity beta}
\end{equation}
with $\mathcal{R}_{d,\beta}$ defined in Eq.~(\ref{eq. Riesz beta}). Therefore,
\begin{equation}
\begin{aligned}
\frac{C(r,t)}{C_{\rm st}(r)}
&=\Phi_{d,\beta}\!\left(
\frac{\mathcal D_\beta^{(d)}t}{r^\beta}\right),\\
\Phi_{d,\beta}(\tau)&\equiv
\frac{H_{d,\beta}(\tau)}{H_{d,\beta}(\infty)}\,.
\end{aligned}
    \label{eq. corr vs corrst beta}
\end{equation}
For $r\ll\xi(t)$, $\Phi_{d,\beta}\to1$ and the correlations have already reached the stationary profile, whereas for $r\gg\xi(t)$ they have not. The crossover therefore occurs at $r\sim\xi(t)$. Eq.~(\ref{eq. xi beta}) gives $\xi(t)\sim t^{1/\sigma}$ in the long-range-dominated regime and $\xi(t)\sim t^{1/2}$ in the diffusive regime.

For $\sigma>2$, where $\beta=2$, Eq.~(\ref{eq. H representation beta}) can be evaluated explicitly. Using the Gaussian propagator,
\begin{equation}
    C(r,t)
    \simeq
    \frac{v_{\rm st}}
    {4\pi^{d/2}B_{\sigma}^{(d)}}\,
    r^{2-d}\,
    \Gamma
    \left(
    \frac{d-2}{2},
    \frac{r^2}{4B_{\sigma}^{(d)}t}
    \right),
    \label{eq. short-range correlation td}
\end{equation}
where $\Gamma(a,x)$ is the upper incomplete Gamma function. This extends the known result for the purely short-range dynamics~\cite{HenkelStoimenov2026}. Consequently,
\begin{equation}
\begin{aligned}
\frac{C(r,t)}{C_{\rm st}(r)}
&=\frac{\Gamma\bigl(\frac{d-2}{2},
\frac{r^2}{4B_\sigma^{(d)}t}\bigr)}
{\Gamma\bigl(\frac{d-2}{2}\bigr)},\\
&\hspace{0.5em}\sigma>2\,.
\end{aligned}
    \label{eq. short range ratio explicit}
\end{equation}
This is the explicit diffusive realization of the unified scaling form in Eq.~(\ref{eq. corr vs corrst beta}).

\bibliography{references}

\end{document}